# Computational Properties of Metaquerying Problems [*]

F. Angiulli

ISI-CNR, c/o DEIS,

Università della Calabria,

87036 Rende (CS), Italy

angiulli@isi.cs.cnr.it

R. Ben-Eliyahu-Zohary

CSE, Ben Gurion University of the Negev,

Ber-Sheeva 84105, Israel

rachel@bgumail.bgu.ac.il

G. Ianni[†]

DEIS, Università della Calabria,

87036 Rende (CS), Italy

ianni@deis.unical.it

L. Palopoli

DIMET, Università di Reggio Calabria,

89100 Reggio Calabria (RC), Italy

palopoli@ing.unirc.it

**Abstract**

Metaquerying is a datamining technology by which hidden dependencies among several database relations can be discovered. This tool has already been successfully applied to several real-world applications. Recent papers provide only preliminary results about the complexity of metaquerying. In this paper we define several variants of metaquerying that encompass, as far as we know, all variants defined in the literature. We study both the combined complexity and the data complexity of these variants. We show that, under the combined complexity measure, metaquerying is generally intractable (unless $P=NP$), lying sometimes quite high in the complexity hierarchies (as high as $NP^{PP}$), depending on the characteristics of the plausibility index. However, we are able to single out some tractable and interesting metaquerying cases (whose combined complexity is LOGCFL-complete). As for the data complexity of metaquerying, we prove that, in general, this is in $TC^0$, but lies within $AC^0$ in some simpler cases. Finally, we discuss implementation of metaqueries, by providing algorithms to answer them.

## 1 Introduction

Companies and organizations often posses information resources and databases containing a vast amount of data waiting to prove useful one day. The expanding datamining research area is supposed to provide tools for

---

the discovery of valuable knowledge in these huge data resources [14].

Metaquerying [22] is a promising approach for datamining in relational and deductive databases. Metaqueries serve as a generic description of a class of patterns a user is willing to discover. Unlike many other mining tools, patterns discovered using metaqueries can link information from several tables in databases. These patterns are all relational, while most machine-learning systems can only learn propositional patterns and work on a single relation. Metaqueries can be specified by human experts or alternatively, they can be automatically generated from the database schema.

Intuitively, a metaquery has the form

$$T \leftarrow L_1, ..., L_m \qquad (1)$$

where $T$ and $L_i$ are literal schemes $Q(Y_1, ..., Y_n)$ and Q is either an ordinary predicate name or a predicate variable. In this latter case, $Q(Y_1, ..., Y_n)$ can be instantiated to an atom with predicate symbol denoting a relation in the database. An answer to a metaquery is an (ordinary) rule obtained by consistently substituting second order predicates with relation names.

Shen $et al.$ [21] are, to best of our knowledge, the first who have presented a framework that uses metaqueries to integrate inductive learning methods with deductive database technology.

As an example (taken from [22]), let $P,Q$, and $R$ be predicate variables and **DB** be a database, then the metaquery

$$R(X, Z) \leftarrow P(X, Y), Q(Y, Z).$$

specifies that the patterns to be discovered are relationships of the form

$$r(X, Z) \leftarrow p(X, Y), q(Y, Z).$$

where $p$, $q$, and $r$ are relations from **DB**. For instance, for an appropriate database **DB**, one possible result of this metaquery could be a rule:

$$speaks(X, Z) \leftarrow citizen(X, Y),\ language(Y, Z). \qquad (2)$$

A rule which serves as an answer to a metaquery is usually accompanied by indices, that indicate its "plausibility degree". In [8], for example, each rule in the answer is supplied with *support* and *confidence*. The *support* indicates how frequently the body of the rule is satisfied, and the *confidence* measures what fraction of the tuples that satisfy the body, also satisfy the head. A confidence of 0.93, for instance, for the rule (2) means that out of all pairs $(X, Z)$ that satisfy the body of the rule, 93% also satisfy the head. Admissibility thresholds for support and confidence is usually provided by the user. Support and confidence have been also defined for other datamining techniques, such as association rules [3].

Similar to the case of association rules, the plausibility indices are used with two major purposes:

1. to avoid presenting negligible information to the user,

2. to cut off the search space by early detection of plausibility values.

Formal definition of metaqueries and indices are given in the next section.

Metaquerying was implemented in some datamining systems [12, 11, 22] and was arguably very useful in knowledge discovery [15, 20, 24, 22]. However, despite the practical success of metaqueries, their theoretical foundations appear to be quite unclear. A careful reading of research manuscripts on the subject reveals that the semantics is not always well defined. In addition, except a very preliminary *NP*-hardness result [8], a thorough analysis of the involved complexities was never performed.

This paper reports a study of some of the theoretical aspects about metaquerying. First, in Section 2, we provide formal definitions for the syntax and semantics of metaqueries. We define several types of metaqueries and indices that encompass, as far as we know, all variants defined in the literature.

In Section 3 we study both the combined complexity and the data complexity of all the metaqueries variants that we define. We show that, under the combined complexity measure, metaquerying is intractable in general (unless *P=NP*), but we are able to single out a tractable interesting metaquerying case (whose combined complexity is *LOGCFL*-complete). The tractable class is a subset of metaquerying problems which we call acyclic metaqueries. Acyclicity of metaqueries is a generalization of the analogous concept defined for conjunctive queries [7].

As for the data complexity of metaquerying, we prove that, in general, this is in $TC^0$, but lies within $AC^0$ in some interesting cases.

In Section 4 we discuss algorithms to implement metaquerying in a reasonably efficient manner.

Finally, in section 5, we draw some conclusions and report a summary table containing all the complexity results we prove throughout the paper.

## 2 Metaquerying

### 2.1 Syntax and semantics

Let $U$ be a countable domain of constants. A database **DB** is $(D, R_1, ..., R_n)$ where $D \subseteq U$ is finite, and each $R_i$ is a relation of arity $a(R_i)$ such that $R_i \subseteq D^{a(R_i)}$.

As stated above, a metaquery **MQ** is a second-order template describing a pattern to be discovered [22]. Such a template has the form

$$T \leftarrow L_1, ..., L_m \tag{3}$$

where $T$ and $L_i$ are *literal schemes*. Each literal scheme $T$ or $L_i$ has the form $Q(Y_1, ..., Y_n)$ where $Q$ is either a *predicate (second order) variable* or a relation symbol, and each $Y_j$ ($1 \leq j \leq n$) is an ordinary (first order) variable. If $Q$ is a predicate variable, then $Q(Y_1, ..., Y_n)$ is called a *relation pattern* of arity $n$, otherwise it is called an *atom* of arity $n$. The right-hand-side $L_1,...,L_m$ is called the *body* of the metaquery, while $T$ is called the *head* of the metaquery. A metaquery is called *pure* if each two relation patterns with the same predicate variable has the same arity.

Intuitively, given a database instance **DB**, answering a metaquery **MQ** on **DB** amounts to finding all substitutions $\sigma$ of relation patterns appearing in **MQ** by atoms having as predicate names relations in **DB**, such that the Horn rule $\sigma(\mathbf{MQ})$ (obtained by applying $\sigma$ to **MQ**) encodes a dependency between the atoms in its head and body, which rule holds in **DB** with a certain degree of plausibility. The plausibility is defined in terms of *indices* which we will formally define shortly.

Let **MQ** be a metaquery and **DB** a database. Let $pv(\mathbf{MQ})$, $ls(\mathbf{MQ})$, and $rep(\mathbf{MQ})$ denote the set of predicate variables, the set of literal schemes, and the set of relation patterns occurring in **MQ**, respectively (note that $rep(\mathbf{MQ}) \subseteq ls(\mathbf{MQ})$). Moreover, let $rel(\mathbf{DB})$ denote the set of relation names of **DB** and $ato(\mathbf{DB})$ denote the set of all the atoms of the form $p(T_1, \ldots, T_k)$ where $p \in rel(\mathbf{DB})$, $k$ is the arity of $p$, and each $T_i$ is an ordinary variable.

Semantics is defined via types of *metaquery instantiations*: an instantiation type specifies how relation patterns can be instantiated, turning a metaquery to an ordinary Horn rule over the given database. Next, we define three different types of metaquery instantiations, that we call *type-0, type-1* and *type-2*, respectively. In the literature, metaquerying semantics has not been always precisely defined, even if a kind of type-2 semantics is usually assumed (see, e.g.,[21]).

**Definition 2.1** Let **MQ** be a metaquery and **DB** a database. An *instantiation* (on **MQ** and **DB**) is a mapping $\sigma : rep(\mathbf{MQ}) \to ato(\mathbf{DB})$, whose restriction $\sigma' : pv(\mathbf{MQ}) \to rel(\mathbf{DB})$ is functional.

The condition above says that predicate names of **MQ** are *consistently* substituted with relation names from **DB**.

In general, given a set $S$ of literal schemes and an instantiation $\sigma$, $\sigma(S)$ will denote a set of atoms generated from $S$ by applying $\sigma$ to each relation pattern belonging to $S$. Let **DB** be a database, and **MQ** be a metaquery. In the following, $\sigma(\mathbf{MQ})$ will denote the Horn rule generated by applying $\sigma$ to each relation pattern of **MQ**. Whenever we want to remark that $\sigma$ is from $rep(\mathbf{MQ})$ to $ato(\mathbf{DB})$, we will employ the notation $\sigma_{\mathbf{DB}}^{\mathbf{MQ}}$.

**Definition 2.2** Let **MQ** be a pure metaquery and **DB** a database. An instantiation $\sigma_{\mathbf{DB}}^{\mathbf{MQ}}$ is *type-0* if for any relation pattern $L$ and atom $A$, $\sigma(L) = A$ implies that $L$ and $A$ have the same list of arguments.

That is, under type-0 semantics, each predicate variable is always matched to a relation with the same arity and ordinary variables are left untouched. As an example, consider the database $\mathbf{DB}_1$ shown in Figure 1 and the metaquery

$$R(X,Z) \leftarrow P(X,Y), Q(Y,Z) \tag{4}$$

A possible type-0 instantiation for **MQ** is

$$\sigma = \{\langle R(X,Z), \mathrm{UsPT}(X,Z)\rangle, \langle P(X,Y), \mathrm{UsCa}(X,Y)\rangle, \langle Q(Y,Z), \mathrm{CaTe}(Y,Z)\rangle\}$$

which yields the following Horn rule when applied to **MQ**:

$$\mathrm{UsPT}(X,Z) \leftarrow \mathrm{UsCa}(X,Y), \mathrm{CaTe}(Y,Z)$$

**Definition 2.3** Let **MQ** be a pure metaquery and **DB** a database. An instantiation $\sigma_{\mathbf{DB}}^{\mathbf{MQ}}$ is *type-1* if for any relation pattern $L$ and atom $A$, $\sigma(L) = A$ implies that the arguments of $A$ are obtained from arguments of $L$ by permutation.

With type-1 instantiations, variable ordering within relation patterns "does not matter". As an example, under this semantics, from metaquery (4) and $\mathbf{DB}_1$, among others, the following Horn rules could be obtained:

$$\mathrm{UsPT}(X,Z) \leftarrow \mathrm{UsCa}(X,Y), \mathrm{CaTe}(Y,Z)$$
$$\mathrm{UsPT}(X,Z) \leftarrow \mathrm{UsCa}(Y,X), \mathrm{CaTe}(Y,Z)$$

The third type of instantiation takes a step further by allowing a relation pattern of arity $k$ to be matched with an atom of arity $k'$, with $k' \geq k$, padding "remaining" arguments to free variables:

**Definition 2.4** Let **MQ** be a metaquery and **DB** a database. An instantiation $\sigma_{\mathbf{DB}}^{\mathbf{MQ}}$ is *type-2* if for any relation pattern $L$ and atom $A$, $\sigma(L) = A$ implies the following:

- the arity $k'$ of $A$ is greater-than or equal-to the arity of $L$;
- $k$ of the arguments of $A$ coincide with the $k$ arguments of $L$, possibly occurring in different positions;
- the remaining $k' - k$ arguments of $A$ are variables not occurring elsewhere in the instantiated rule.

With type-2 instantiations we can express interesting patterns ignoring how many extra attributes a physical relation may have. Should the relation *UsPT* be defined with an additional attribute, as in Figure 2, the metaquery (4) can be instantiated, using a type-2 instantiation, to

$$\mathrm{UsPT}(X,Z,T) \leftarrow \mathrm{UsCa}(Y,X), \mathrm{CaTe}(Y,Z)$$

UsCa

| User | Carrier |
|---|---|
| John K. | Omnitel |
| John K. | Tim |
| Anastasia A. | Omnitel |

CaTe

| Carrier | Technology |
|---|---|
| Tim | ETACS |
| Tim | GSM 900 |
| Tim | GSM 1800 |
| Omnitel | GSM 900 |
| Omnitel | GSM 1800 |
| Wind | GSM 1800 |

UsPt

| User | Phone Type |
|---|---|
| John K. | GSM 900 |
| John K. | GSM 1800 |
| Anastasia A. | GSM 900 |

Figure 1: The relations UsCa, CaTe, and UsPT of $\mathbf{DB}_1$

| UsPt | | |
|---|---|---|
| User | Phone Type | Model |
| John K. | GSM 900 | Nokia 6150 |
| John K. | GSM 1800 | Nokia 6150 |
| Anastasia A. | GSM 900 | Bosch 607 |

Figure 2: The new relation *UsPT*

Note that a type-0 instantiation is a type-1 instantiation where the chosen permutation of relations's attributes is the identity, whereas a type-1 instantiation is a type-2 instantiation, where the arity of the atoms matches the arity of the relation patterns they are substituted for.

Note, moreover, that type-2 instantiations may apply to any metaquery, while type-0 and type-1 instantiations require pure metaqueries.

## 2.2  Plausibility Indices

In datamining applications, one is generally interested in discovering plausible patterns of data that represent significant knowledge in the analyzed data sets. In other words, it is not typically required for a discovered rule to entirely characterize the given data set, but a significant subset thereof. This idea is embedded in the usage of *plausibility indices*. In the literature, several definitions of plausibility indices are found, like, e.g., *support*, *confidence*, *base* and *strength* [3, 8, 21] (in fact, "support" is similar to "base" and "confidence" is similar to "strength" [8]). In the analysis that follows, we shall use support, confidence and another useful index that we call *cover*. We next provide formal definitions of the indices that we use.

For any set $F$, let $|F|$ denote its size. In the following, we assume the reader is familiar with basic notions regarding relational algebra and Datalog (see [28] for an excellent source of material about these subjects). Unless otherwise specified, a predicate name $p$ will denote its corresponding underlying relation as well. In particular, an atom $p(\mathbf{X})$ will denote the corresponding database relation, where the list of arguments $\mathbf{X}$ is used, as in Datalog, to positionally refer to $p$'s columns. For a set of atoms $R$, $att(R)$ is the set of all the variables (attribute) of all the atoms in $R$, and $\mathbf{J}(R)$ is the natural join of all the relations corresponding to atoms in $R$.

**Definition 2.5** A plausibility index, or index in short, is a function which maps a database instance $\mathbf{DB}$ and a Horn rule $h(\mathbf{X}) \leftarrow b_1(\mathbf{X}_1), ..., b_n(\mathbf{X}_n)$ (defined over $\mathbf{DB}$) to a rational number within $[0, 1]$.

Given an index $I$, a database $\mathbf{DB}$, and a rule $r$ defined over $\mathbf{DB}$, unless confusion may arise, we will employ the notation $I(r)$ as a shortcut for $I(r, \mathbf{DB})$.

**Definition 2.6** Let $R$ and $S$ be two sets of atoms. Then the fraction of $R$ in $S$, denoted $R \uparrow S$, is

$$\frac{|\pi_{att(R)}(\mathbf{J}(R) \bowtie \mathbf{J}(S))|}{|\mathbf{J}(R)|}.$$

In particular, whenever $|\pi_{att(R)}(\mathbf{J}(R) \bowtie \mathbf{J}(S))| = 0$, $R \uparrow S$ is defined equal to 0.

**Definition 2.7** Let **DB** be a database, and $r$ be a Horn rule defined on **DB**. Let $h(r)$ and $b(r)$ denote the sets of atoms occurring in the head and in the body of $r$, respectively. Then

- the *confidence* of $r$ on **DB** is $cnf(r) = b(r) \uparrow h(r)$,

- the *cover* of $r$ on **DB** is $cvr(r) = h(r) \uparrow b(r)$,

- the *support* of $r$ on **DB** is $sup(r) = \max_{a \in b(r)}(\{a\} \uparrow b(r))$.

Intuitively, $sup(r)$ measures how much the body (or part of it) of an instantiation contains satisfying tuples. When an instantiation scores an high support, a pattern search algorithm may conclude that it is worth to further consider such an instantiation, because there is at least one relation with a high percentage of its tuples satisfying the instantiated body.

When an instantiation scores a high confidence, we can conclude that a high percentage of the assignments which satisfy the body also satisfy the head relation. Hence, confidence tells how much valid the rule is over the given database. Given a rule $r$, the indices $cnf(r)$ and $sup(r)$ are equivalent to *confidence* and *support* defined in [8]. In [8], there is also a discussion on the motivations underlying various definition of support and confidence.

To conclude, cover tells which is the percentage of implied tuples belonging to the head relation. This latter index that we define here, is useful in those application where it is necessary to decide if it is worth to store the head relation or to compute it in the form of a reasonably matching view, such as in reengineering applications.

As an example, consider the metaquery

$$I(X) \leftarrow O(X)$$

the (type-2) instantiation (on $\mathbf{DB}_1$)

$$\text{UsCa}(X, Z) \leftarrow \text{UsPt}(X, H)$$

scores a cover of 1, which tells that the content of the first attribute of UsPt completely encodes the content of the first attribute of UsCa.

In the sequel, the set of plausibility indices $\{cnf, cvr, sup\}$ will be denoted **I**.

## 3 Complexity of metaquerying

In this section we present several results regarding the complexity of metaquerying. We assume the reader is familiar with basic notions regarding complexity classes and, in particular, the definition of the polynomial hierarchy [26]. Next, we recall the definition of the classes $AC^0$, $TC^0$, $\#AC^0$, $GapAC^0$, $PAC^0$, and $LOGCFL$, which will be used in the sequel.

## 3.1 Preliminaries

**Definition 3.1** Let $\mathcal{C}$ be a complexity class, then $NP^{\mathcal{C}}$ is the class of problems decidable in $NP$ using an oracle solving a problem in $\mathcal{C}$ (see [31]).

**Definition 3.2** A *conjunctive query* is a set of atoms $\{r_1(\mathbf{X}_1), \ldots, r_n(\mathbf{X}_n)\}$, where $\mathbf{X}_1, \ldots, \mathbf{X}_n$ are lists of variables and/or constants. Let **DB** be a database instance. The problem of satisfying a conjunctive query (Boolean Conjunctive Query satisfaction problem, or BCQ) is the problem of deciding if there exists a substitution $\rho$ for variables in $X_i, 1 \leq i \leq n$ such that, for each $i$, $1 \leq i \leq n$, $r_i(\rho(\mathbf{X}_i)) \in \mathbf{DB}$. The set $\{r_i(\rho(\mathbf{X}_i)), 1 \leq i \leq n\} \equiv \{r_i(\mathbf{x}_i), 1 \leq i \leq n\}$ is called *ground instance* of $\{r_1(\mathbf{X}_1), \ldots, r_n(\mathbf{X}_n)\}$.

The BCQ problem was proved to be *NP*-complete in [10].

**Definition 3.3** MAJORITY gates are unbounded fan-in gates (with binary input and output) that output 1 if and only if more than half of their inputs are non-zero.

**Definition 3.4** A family $\{C_i\}$ of boolean circuits, such that $C_i$ accepts strings of size $i$, is uniform if there exists a Turing machine $T$ which, on input $i$ produces the circuit $C_i$. $\{C_i\}$ is said to be *logspace uniform* if $T$ carries out its work using $O(\log i)$ space. Define $AC^0$ (resp. $TC^0$) as the class of decision problems solved by logspace uniform families of circuits of polynomial size and constant depth, with AND, OR, and NOT (resp. MAJORITY and NOT) gates with unbounded fan-in [2, 6, 23].

In the following, unless otherwise specified, we will assume to deal with logspace uniform families of circuits. The class $TC^0$ is of special interest, since it characterizes the computational complexity of such important operations as multiplication, division, and sorting, as well as being a computational model for neural nets. $TC^0$ was also characterized as being the class of languages that arises in several ways from counting the number of accepting subtrees of $AC^0$ circuits [2].

**Definition 3.5** For any $k > 0$, $\#AC^0_k$ is the class of functions $f : \{0,1\}^* \to \mathbf{N}$ computed by depth $k$, polynomial size uniform families of circuits with $+, \times$-gates (the usual arithmetic sum and product in $\mathbf{N}$) having unbounded fan-in, where each value incoming into the circuit can be either constant (where the allowed constant values are 1 and 0) or being an input value in the form $x_i$ or $1 - x_i$ (where the allowed input values are 1 and 0). Let $\#AC^0 = \bigcup_{k>0} \#AC^0_k$ [2].

Thus, $\#AC^0$ circuits accept the values 1 and 0 as inputs, but they are considered as natural numbers.

**Definition 3.6** $GapAC^0$ is the class of all functions $f : \{0,1\}^* \to \mathbf{N}$ that can be expressed as the difference of two functions in $\#AC^0$ [2, 4].

**Definition 3.7** $PAC^0$ is the class of languages $\{A \mid \exists f \in GapAC^0, x \in A \iff f(x) > 0\}$ [2].

**Proposition 3.8** Under logspace uniformity, $PAC^0 = TC^0$ [2, 4].

**Definition 3.9** *LOGCFL* coincides with the class of decision problems logspace-reducible to a context free language [23].

**Definition 3.10** Let $H(\mathbf{x}, \mathbf{y})$ be a formula with two free variable lists $\mathbf{x}$ and $\mathbf{y}$ and let $k$ be a natural number. The *counting quantifier* $\mathbf{C}$ is defined as

$$\overset{k}{\underset{\mathbf{y}}{\mathbf{C}}} H(\mathbf{x}, \mathbf{y}) \equiv |\{\mathbf{y} : H(\mathbf{x}, \mathbf{y}) \text{ is true}\}| \geq k$$

[33].

As an example

$$\overset{k}{\underset{\mathbf{y}}{\mathbf{C}}} [\mathbf{y} \text{ encodes an hamiltonian path on the graph } \mathbf{x}]$$

is true iff the number of Hamiltonian paths on $\mathbf{x}$ is at least $k$. A polynomial-bounded version of the counting quantifier is defined as follows.

**Definition 3.11** Let $\mathcal{L}$ be a class of languages. A language $L'$ belongs to the class $C\mathcal{L}$ iff there exists a language $L \in \mathcal{L}$, a polynomial time computable function $f$, and a polynomial $p$ such that

$$x \in L' \leftrightarrow \overset{f(x)}{\underset{\substack{y \\ |y| \leq p(|x|)}}{\mathbf{C}}} [(x, y) \in L]$$

The polynomial-bounded counting quantifier embeds the polynomial-bounded existential quantifier as follows: A language $L'$ belongs to the class $\exists \mathcal{L}$ iff there exists a language $L \in \mathcal{L}$, and a polynomial $p$ such that

$$x \in L' \leftrightarrow \overset{1}{\underset{\substack{y \\ |y| \leq p(|x|)}}{\mathbf{C}}} [(x, y) \in L]$$

**Definition 3.12** The language $\exists \mathbf{C}$-SAT, includes all tuples $\langle F(\mathbf{x}_1, \mathbf{x}_2), m \rangle$ (where $\mathbf{x}_1$ and $\mathbf{x}_2$ are lists of propositional variables) such that $F$ is a boolean expression in conjunctive normal form and

$$\overset{1}{\underset{\mathbf{x}_1}{\mathbf{C}}} \overset{m}{\underset{\mathbf{x}_2}{\mathbf{C}}} F(\mathbf{x}_1, \mathbf{x}_2) \text{ is true}$$

**Theorem 3.13** $\exists \mathbf{C}$-*SAT and* $\exists \mathbf{C}$-*3SAT are* $\exists CP$-*complete* [33].

Let $PP$ be the set of languages $\mathcal{L}$ recognized by a nondeterministic Turing Machine $M$ as follows: $x \in \mathcal{L}$ iff at least one half of the computation paths of $M(x)$ ends in an accepting state. It is known that $\exists CP$ coincides with $NP^{PP}$ [27]. $\exists CP$ belongs to the so called polynomial counting hierarchy: it is contained in $PSPACE$, and (may be properly) contains $\Sigma_2^P$ and $CP$ [33].

**Definition 3.14** A *counting Turing Machine (CTM) CM* is an ordinary nondeterministic Turing Machine whose output, on the input $x$, is the number of accepting computations of $CM(x)$. The time complexity of a CTM is defined to be $f(n)$ if the longest accepting computation associated with the set of all inputs of size $n$ takes $f(n)$ steps (see [19]).

**Definition 3.15** Define $\#P$ to be the set of all functions that are computable by polynomial-time CTMs.

**Definition 3.16** A problem $X$ is $\#P$-hard if there are polynomial-time Turing reductions from all problems in $\#P$ to $X$. If $X$ is $\#P$-hard and $X \in \#P$, then $X$ is said to be $\#P$-complete (see [30]).

**Definition 3.17** A *parsimonious transformation* [19] is a polynomial transformation $f$ from a problem $X$ to a problem $Y$ such that, if $\#(X,x)$ is defined to be the number of solutions that instance $x$ has in problem X, then $\#(X,x) = \#(Y,f(x))$.

## 3.2 Complexity Measures

As for the case of ordinary queries, complexity of metaqueries can be defined according to two complexity measures: *combined complexity* and *data complexity* [32]. Complexity measures are defined next.

Let $T \in \{0,1,2\}$ be an instantiation type and let $I$ be a plausibility index. Let **DB** denote a database instance, **MQ** a metaquery, and $k$ a rational threshold value such that $0 \leq k < 1$. Then:

1. The *combined* complexity of $\langle \mathbf{DB}, \mathbf{MQ}, I, k, T \rangle$ is the complexity, measured in the size of **DB**, **MQ** and $k$, of deciding if there exists a type-$T$ instantiation $\sigma_{\mathbf{DB}}^{\mathbf{MQ}}$ such that $I(\sigma(\mathbf{MQ})) > k$.

2. Assuming that a database schema $DS = (D, R_1, ..., R_n)$ has been fixed in advance, the *data* complexity of the metaquery problem $\langle \mathbf{DB}, \mathbf{MQ}, I, k, T \rangle$ is the complexity, measured in the size of **DB**, of deciding if there exists a type-$T$ instantiation $\sigma_{\mathbf{DB}}^{\mathbf{MQ}}$ such that $I(\sigma(\mathbf{MQ})) > k$, where **DB** is a database with schema $DS$.

In the literature it is usually assumed that one looks for rules that satisfy some metaquery and have two plausibility indices, usually *support* and *confidence*, above some given threshold [8, 21]. Here we split the metaquery problem so that it relates to one plausibility index at a time. The rationale is that, this allows us to single out more precisely complexity sources and, at the same time, complexity measures for problems involving more than one index can be obtained fairly easily from metaquerying problems involving only one index.

## 3.3 Combined Complexity

It immediately follows from the proof of Theorem 2.2 in [8] that the combined complexities of $\langle \mathbf{DB}, \mathbf{MQ}, sup, 0, 1 \rangle$ and $\langle \mathbf{DB}, \mathbf{MQ}, cnf, 0, 1 \rangle$ are NP-hard. In this section, we generalize these results by stating the combined complexities of metaquerying for various indices and instantiation types.

A first characterization of the problem is given by the following result.

**Proposition 3.18** *Let $I$ be an index, and $\mathcal{C}$ be the complexity of deciding the following question: "Given a Horn rule $r$, a database $\mathbf{DB}$ and a finitely represented rational value $k \in [0,1)$, is $I(r) > k$ over $\mathbf{DB}$?". Then, for any instantiation type $T$, the complexity of $\langle \mathbf{DB}, \mathbf{MQ}, I, k, T \rangle$ is in $NP^{\mathcal{C}}$.*

**Proof.** Simply note that the evaluation of an instance of $\langle \mathbf{DB}, \mathbf{MQ}, I, k, T \rangle$ can be done by guessing an instantiation $\sigma_{\mathbf{DB}}^{\mathbf{MQ}}$, and then calling a $\mathcal{C}$ oracle to decide if $I(\sigma(\mathbf{MQ})) > k$. □

In many cases, we are able to state tighter bounds than those implied by Proposition 3.18. We begin by analyzing the cases when the threshold value $k$ is set to 0. This kind of metaquerying problems have a limited practical usefulness. However, they provide interesting lower bounds for the complexity of metaquerying problems involving thresholds larger than 0.

**Definition 3.19** *Let $\mathbf{DB}$ be a database, and $r \equiv b_1(\mathbf{X}_1) \leftarrow b_2(\mathbf{X}_2), ..., b_n(\mathbf{X}_n)$ be a Horn rule defined over $\mathbf{DB}$. Let $I$ be an index, and let $A_r = \{b_1(\mathbf{X}_1) \leftarrow b_2(\mathbf{X}_2), ..., b_n(\mathbf{X}_n)\}$ the set of atoms occurring in $r$. A certifying set for $I$ is a set $S \subseteq A_r$ such that there exists a ground instance $s$ of $S$ which is satisfied in $\mathbf{DB}$ iff $I(r) > 0$.*

**Proposition 3.20** *Consider a rule $r \equiv h(\mathbf{X}) \leftarrow b_1(\mathbf{X}_1), ..., b_n(\mathbf{X}_n)$. Then, $S_{\text{cvr}} \equiv \{h(\mathbf{X}), b_1(\mathbf{X}_1), ..., b_n(\mathbf{X}_n)\}$, $S_{\text{sup}} \equiv \{b_1(\mathbf{X}_1), ..., b_n(\mathbf{X}_n)\}$ and $S_{\text{cnf}} \equiv S_{\text{cvr}}$ are certifying sets for $\text{cvr}, \text{sup}$ and $\text{cnf}$, respectively.*

**Proof.** As for cover, consider a rule $r \equiv h(\mathbf{X}) \leftarrow b_2(\mathbf{X}_1), ..., b_n(\mathbf{X}_n)$ defined over a database $\mathbf{DB}$. Since $cvr(r)$ is defined as

$$\frac{|\pi_{\mathbf{X}}\left(h(\mathbf{X}) \bowtie b_1(\mathbf{X}_1) \bowtie ... \bowtie b_n(\mathbf{X}_n)\right)|}{|h(\mathbf{X})|}$$

It is immediate to note that $h(\mathbf{X}) \bowtie b_1(\mathbf{X}_1) \bowtie ... \bowtie b_n(\mathbf{X}_n)$ is not empty iff there exists a ground instance of $\{h(\mathbf{X}), b_1(\mathbf{X}_1), ..., b_n(\mathbf{X}_n)\}$ which is satisfied with respect to $\mathbf{DB}$ (cf. definition 3.2). Proofs concerning confidence and support are similar. □

**Theorem 3.21** *Let $I \in \mathbf{I}$ be a plausibility index. The combined complexity of $\langle \mathbf{DB}, \mathbf{MQ}, I, 0, T \rangle$ is NP-complete, for any instantiation type $T \in \{0, 1, 2\}$.*

**Proof.** *Hardness.* We use a reduction from the graph 3-COLORING problem [16]; that is, given a undirected graph $G = (V, E)$ and three colors $\{1, 2, 3\}$, is it possible to assign a color to each node of $G$, such that no adjacent nodes have the same color?

The reduction uses a database $\mathbf{DB}_{3col}$ and a metaquery $\mathbf{MQ}_{3col}$. $\mathbf{DB}_{3col}$ includes a single binary relation

$$e = \{(1,2), (1,3), (2,3), (2,1), (3,1), (3,2)\}$$

denoting all the possible ways of correctly coloring two adjacent nodes. In $\mathbf{MQ}_{3col}$ we use ordinary variables $X_u$, one for each node $u \in V$. $\mathbf{MQ}_{3col}$ contains the set of literals $S \equiv \{E(X_{u_1}, X_{u_2}), \ldots, E(X_{u_{2k-1}}, X_{u_{2k}})\}$ where $\{(u_1, u_2), \ldots, (u_{2k-1}, u_{2k})\}$ is the set of edges of $G$, that is, $S$ encodes $G$ as a set of literals. $\mathbf{MQ}_{3col}$ is as follows:

$$E(X_{u_1}, X_{u_2}) \leftarrow E(X_{u_1}, X_{u_2}), \ldots, E(X_{u_{2k-1}}, X_{u_{2k}}).$$

**Claim 3.22** *$G$ has a 3-coloring iff for any type $T$, there exists a type-$T$ instantiation $\sigma_{\mathbf{DB}_{3col}}^{\mathbf{MQ}_{3col}}$ such that $\sigma(\mathbf{MQ})$ has a ground instance $s$ which is satisfied with respect to $\mathbf{DB}_{3col}$.*

**Proof.** (Claim 3.22) ($\rightarrow$). Suppose that $G$ has a 3-coloring $c : V \mapsto \{1, 2, 3\}$, defined for each $u \in V$, and such that if $(u, v) \in E$ then $c(u) \neq c(v)$. To show this direction, it is enough to consider an instantiation $\sigma$ which maps each literal in the form $E(X, Y)$ to an atom of the form $e(X, Y)$. $\sigma$ is type-0 and, therefore, also type-1 and type-2.

($\leftarrow$). Since $S$ is an encoding of the edges of $G$, $G$ is undirected and $e$ is the only relation in $\mathbf{DB}_{3col}$, $s$ being true over $\mathbf{DB}_{3col}$ implies that $G$ is 3-colorable. This closes the proof of Claim 3.22. □

The hardness part of the proof is completed by noting that by Proposition 3.20 and by the definition of $\mathbf{MQ}_{3col}$, by defining an instantiation $\sigma$ which maps $E$ to $e$, $\sigma(S)$ is a certifying set for *cvr*, *sup*, and *cnf*.

*Membership.* Assume $I = cvr$. The proof consists in providing evidence of the existence of a succinct certificate for the problem [16]. In order to check if $\langle \mathbf{DB}, \mathbf{MQ}, I, 0, T \rangle$ is a YES instance, we guess an instantiation $\sigma_{\mathbf{DB}}^{\mathbf{MQ}}$ of type $T$, and a substitution $\rho$ for ordinary variables and check if $\rho(\sigma(\mathbf{MQ}))$ is true in $\mathbf{DB}$. By definition 3.19, $\rho(\sigma(\mathbf{MQ}))$ is true in $\mathbf{DB}$ iff $I(\sigma(\mathbf{MQ})) > 0$. Proofs concerning support and confidence are similar. □

**Remark.** Our proof outlines a general technique which can be employed to prove analogous results for all those indices $I$ having the following properties:

- Given a rule $r \equiv h(\mathbf{X}_0) \leftarrow b_1(\mathbf{X}_1), \ldots, b_n(\mathbf{X}_n)$ a certifying set $S_I$ can be built in polynomial time, and
- $\mathcal{O}(|S_I|) \geq \mathcal{O}(n^{\frac{1}{k}})$, for some $k \geq 1$.

Next, we consider the combined complexity of metaquerying when the fixed threshold $k$ is such that $0 \leq k < 1$.

**Proposition 3.23** *The combined complexity of $\langle \mathbf{DB}, \mathbf{MQ}, I, k, T \rangle$ is NP-hard for any plausibility index $I \in \mathbf{I}$.*

**Proof.** Immediate from Theorem 3.21. □

**Theorem 3.24** *The combined complexity of $\langle \mathbf{DB}, \mathbf{MQ}, I, k, T \rangle$, with $0 \leq k < 1$, $T \in \{0, 1, 2\}$ and $I \in \{\text{cvr}, \text{sup}\}$ is in NP.*

**Proof.** Consider an instance of $\langle \mathbf{DB}, \mathbf{MQ}, cvr, k, T \rangle$. $I(\sigma(\mathbf{MQ}))$ has the form $\frac{|A|}{|B|}$, where $B$ is the head relation of $\sigma(\mathbf{MQ})$ and $A$ is the join of the body and the head of $\sigma(\mathbf{MQ})$ projected on the attributes of the head relation, for $\sigma_{\mathbf{DB}}^{\mathbf{MQ}}$ a given instantiation of type-$T$. In order to verify if the given instance is a YES instance, it is then sufficient to guess a type-$T$ instantiation $\sigma_{\mathbf{DB}}^{\mathbf{MQ}}$ and verify that, over $\mathbf{DB}$, $|A| > \lfloor k|B| \rfloor$. $k|B|$ can be computed in polynomial time. A succinct certificate for the problem at hand consists of the instantiation $\sigma$, and of $\lfloor k|B| \rfloor + 1$ substitutions $\rho_i$ of ordinary variables occurring in $\sigma(\mathbf{MQ})$ distinct from one another as far as head ordinary variables are concerned, such that $cvr(\sigma(\mathbf{MQ})) > k$ is proved. This can be checked in polynomial time. Checking distinctness of the $\lfloor k|B| \rfloor + 1$ substitutions can obviously be also done in polynomial time.

Consider an instance of $\langle \mathbf{DB}, \mathbf{MQ}, sup, k, T \rangle$. $I(\sigma(\mathbf{MQ}))$ has the form $max_i\{\frac{|A_i|}{|B_i|}\}$, where $B_i$ is a relation in the body of $\sigma(\mathbf{MQ})$ and $A_i$ is the join of the body of $\sigma(\mathbf{MQ})$ projected on the attributes of $B_i$, for $\sigma_{\mathbf{DB}}^{\mathbf{MQ}}$ a given instantiation of type-$T$. The proof follows the same line of reasoning as above, with the difference that, here, the certificate must also include an index $j$ such that $\frac{|A_j|}{|B_j|} > k$. □

As for confidence, it is not difficult to show a *PSPACE* upper bound on its combined complexity and we can deduce, from Theorem 3.21, that the problem is *NP*-hard as well.

A more in-depth analysis shows that the issue of measuring confidence of a given metaquery instantiation turns out to be actually more complex than to measuring the other indices. This is due to the need of computing the *exact* count of tuples satisfying the body of an instantiation, whereas for other indices this is not required. In fact, the problem of deciding if confidence exceeds a given threshold on some instantiation of a metaquery is related to problems where the question concerns *counting* the exact number of solutions of a given instance, as we show below.

Given a boolean formula F (resp. a boolean formula F with at most three literals per clause) in conjunctive normal form, let $\#SAT$ (resp. $\#3SAT$) be the problem of finding the number of satisfying assignments for F.

**Theorem 3.25** $\#SAT$ and $\#3SAT$ are $\#P$-complete [25, 29].

Many problems are known to be $\#P$-complete [29, 13], some of which are the counting counterparts of *NP*-complete problems. However, it is also known that counting solutions for some of the problems in *P* is as hard as counting solutions of *NP*-complete problems [30]. A useful tool to show $\#P$ completeness for counting versions of *NP*-complete problems is that of *parsimonious transformations* (defined in Definition 3.17).

**Proposition 3.26** *Let $Q$ be a conjunctive query $\{r_1(\mathbf{X}_1), \ldots, r_n(\mathbf{X}_n)\}$, where $\mathbf{X}_1, \ldots, \mathbf{X}_n$ are list of variables and/or constants. Let $\#BCQ$ be the problem of counting how many substitutions $\rho$ for variables $\mathbf{X}_i$, $1 \leq i \leq n$, are there such that, for each $i$, $1 \leq i \leq n$, $\rho(r_i(\mathbf{X}_i)) \in \mathbf{DB}$ (cf. Definition 3.2). $\#BCQ$ is $\#P$-complete.*

**Proof.** We show a transformation from 3SAT to BCQ that preserves the number of solutions. Let $F = (x_{11} \vee x_{12} \vee x_{13}) \wedge \cdots \wedge (x_{n1} \vee x_{n2} \vee x_{n3})$ be a 3SAT instance, where the $x_{ij}$'s ($1 \leq i \leq n, 1 \leq j \leq 3$) are (not necessarily distinct) literals. We build a conjunctive query $Q = c_1(X_{11}, X_{12}, X_{13}), \ldots, c_n(X_{n1}, X_{n2}, X_{n3})$ and a database $\mathbf{DB} = c_1, \ldots, c_n$ as follows: each variable $X_{ij}$ of $Q$ is associated with the literal $x_{ij}$ (independently of $x_{ij}$ being negative or not). As for **DB**, each relation $c_i \in$ **DB** is ternary, and the set of constants of **DB** is $U = \{0, 1\}$. Let $cl_i = x_1 \vee x_2 \vee x_3$ be the $i$-th clause of F. We fill $c_i$ with the tuples corresponding to satisfying assignments of $cl_i$, i.e.

$$c_i = U^3 - \{\langle d_1, d_2, d_3 \rangle\}$$

where each $d_j$ ($j \in \{1, 2, 3\}$) is 0 if $x_j$ is positive, and 1 otherwise. Of course, the total number of tuples included into $c_i$ is constant. It is then immediate to see that the number of satisfying assignments for $F$ coincides with the number of satisfying substitutions for $Q$ in **DB**. □

**Theorem 3.27** *The combined complexity of $\langle$**DB**, **MQ**, cnf, $k, T\rangle$ is in $NP^{PP}$.*

**Proof.** We follow the same line of reasoning as in the proof of Theorem 3.24, but here we take advantage of a #BCQ oracle. $cnf(\sigma(\mathbf{MQ}))$ has the form $\frac{|A|}{|B|}$, where $B$ is the join of all body atoms of $\sigma(\mathbf{MQ})$ and $A$ is the join of the body and the head of $\sigma(\mathbf{MQ})$ projected on the attributes of the head relation, for $\sigma_{\mathbf{DB}}^{\mathbf{MQ}}$ a given instantiation of type $T$. In order to verify if the given problem instance is a YES instance, it is then sufficient to guess a type-$T$ instantiation $\sigma$ and verify that, over **DB**, $|A| > \lfloor k|B| \rfloor$. A #BCQ oracle can be queried for values of $|B|$ and $|A|$. A succinct certificate for the problem at hand then consists of the instantiation $\sigma$, and of values $a$ for $|A|$ and $b$ for $|B|$, such that $cnf(\sigma(\mathbf{MQ})) > k$. □

**Theorem 3.28** *The combined complexity of $\langle$**DB**, **MQ**, cnf, $k, 0\rangle$ is $NP^{PP}$ complete.*

**Proof.** (Membership). It is known that #P is as powerful as PP when employed as an oracle, hence $NP^{PP} = NP^{\#P}$ [5]. Membership then immediately follows from Theorem 3.27. (Hardness) We show a reduction from $\exists$**C**-3SAT to $\langle$**DB**, **MQ**, $cnf, k, 0\rangle$.

Suppose an instance of $\exists$**C**-3SAT is given, i.e.:

- a formula $F = \bigwedge_{i=1}^n c_i$, where each $c_i$ is a three-literal clause $l_{i_1} \vee l_{i_2} \vee l_{i_3}$ (with each $l_{i_j} \in \{p_1, \ldots, p_s, q_1, \ldots, q_h, \neg p_1, \ldots, \neg p_s, \neg q_1, \ldots, \neg q_h\}$;

- a partition of variables of $F$ into two sets $\Pi = \{p_1, \ldots p_s\}$ and $\chi = \{q_1, \ldots q_h\}$ and,

- an integer $k'$,

and the question is the following. Is there an assignment for variables $p_1, \ldots, p_s$ such that at least $k'$ assignments for variables $q_1, \ldots, q_h$ are such that $F$ is true?

We build a database $\mathbf{DB}_{csat}$ and a metaquery $\mathbf{MQ}_{csat}$ as follows. The domain of $\mathbf{DB}_{csat}$ is the set of symbols $\{1, 0, l\}$. $\mathbf{DB}_{csat}$ contains:

- a relation $p^a(X, \overline{X}, Y) = \{\langle 1, 0, l \rangle\}$ and a relation $p^b(X, \overline{X}, Y) = \{\langle 0, 1, l \rangle\}$, intended to deal with variables within $\Pi$;

- a relation $q(X, \overline{X}) = \{\langle 1, 0\rangle, \langle 0, 1\rangle\}$, intended to deal with variables within $\chi$;

- a relation

$$c'(L_1, L_2, L_3, C) = \begin{array}{l} \{\langle 1,0,0,1\rangle, \\ \langle 0,1,0,1\rangle, \\ \langle 0,0,1,1\rangle, \\ \langle 1,0,1,1\rangle, \\ \langle 1,1,0,1\rangle, \\ \langle 0,1,1,1\rangle, \\ \langle 1,1,1,1\rangle, \\ \langle 0,0,0,0\rangle\} \end{array}$$

which encodes the possible ways of assigning boolean values to clause literals and, then, to the corresponding clause;

- the relation $c(C_1, \ldots, C_n) = \{\langle 1, \ldots, 1\rangle\}$, intended to select satisfying boolean assignments only.

As for $\mathbf{MQ}_{csat}$, we introduce, for each boolean variable $p_i$ (resp. $q_i$) of $\Pi$ (resp. $\chi$), two variables $P_i$ and $\overline{P}_i$ (resp. $Q_i$ and $\overline{Q}_i$). $\mathbf{MQ}_{csat}$ is as follows:

$$c(C_1, \ldots, C_n) \leftarrow P'_1(P_1, \overline{P}_1, Y), \ldots, P'_s(P_s, \overline{P}_s, Y), q(Q_1, \overline{Q}_1), \ldots, q(Q_h, \overline{Q}_h),$$
$$c'(L_{1_1}, L_{1_2}, L_{1_3}, C_1), \ldots, c'(L_{n_1}, L_{n_2}, L_{n_3}, C_n)$$

where each $L_{i_j}$ ( $1 \leq i \leq n, 1 \leq j \leq 3$), will be

- either $P_y$ if $l_{i_j} = p_y$ and $p_y$ belongs to $\Pi$, or $\overline{P}_y$ if $l_{i_j} = \neg p_y$ and $p_y$ belongs to $\Pi$.

- either $Q_y$ if $l_{i_j} = q_y$ and $q_y$ belongs to $\chi$, or $\overline{Q}_y$ if $l_{i_j} = \neg q_y$ and $q_y$ belongs to $\chi$.

Finally, we set $k$ equal to $\frac{k'-1}{2^h}$.

As an example, suppose $F = (a \vee b \vee e) \wedge (\neg a \vee e \vee d)$, $\Pi = \{a, b\}$ and $\chi = \{d, e\}$. In this case, $\mathbf{MQ}_{csat}$ is

$$c(C_1, C_2) \leftarrow A'(A, \overline{A}, Y), B'(B, \overline{B}, Y), q(D, \overline{D}), q(E, \overline{E}), c'(A, B, E, C_1), c'(\overline{A}, E, D, C_2)$$

Roughly speaking, this reduction works as follows: the generic relation $c'(\_, \_, \_, C_i)$ expresses all the possible value assignment to variables of a clause $c_i$, and the corresponding patterns within $\mathbf{MQ}_{csat}$ encode the structure of $F$; the predicate variables $P'_i$ guess a value, either *true* or *false* for each literal within $\Pi$, whereas, through the atoms with $q$ as the predicate symbol, we encode, in order to *count* them, all the possible assignments of literals within $\chi$. In order to exclude additions to the confidence value determined by instantiations which map some predicate variable $P'_j$ to $q$, $p^a$ and $p^b$ both have arity equal to three. The join of the body atoms, for a given instantiation $\sigma$, "computes" the possible assignments of the variables $Q_i$'s for a fixed configuration of the variables $P_i$'s; when the body of $\sigma(\mathbf{MQ}_{csat})$ is joined with the head atom, the result set captures those assignments on the $Q_i$'s which make $F$ true. Confidence will exceed $k = \frac{k'-1}{2^h}$ iff there are at least $k'$ assignments for $Q_i$'s which make $F$ true.

More formally, suppose that $\langle F, k', \Pi, \chi \rangle$ is a YES instance for $\exists$**C-3SAT**. Consider a truth assignment $\rho$ over variables $p_j$ $(1 \leq j \leq s)$ which makes $\langle F, k', \Pi, \chi \rangle$ a YES instance. An instantiation $\sigma_{\mathbf{MQ}_{3col}}^{\mathbf{DB}_{3col}}$ which makes $\mathit{cnf}(\sigma(\mathbf{MQ}_{3col})) > \frac{k'-1}{2^h}$ can be built by setting, for each $p_j$ $(1 \leq j \leq s)$, $P'_j$ to $p^a$, if $\rho(p_j) = \mathit{true}$ and setting $P'_j$ to $p^b$ otherwise. The join $J$ of body atoms of $\sigma(\mathbf{MQ}_{csat})$ contains precisely $2^h$ tuples. Moreover, if there are at least $k'$ satisfying assignments for variables within $\chi$, joining the head relations with $J$ will result in a relation $J_h$ (which "selects" only those tuples of $J$ such that each $C_i$ variable is set to 1, i.e. the clause $c_i$ is satisfied) containing at least $k'$ tuples. Hence $\mathit{cnf}(\sigma(\mathbf{MQ}_{csat})) > k$.

Conversely, suppose that we have an instantiation $\sigma_{\mathbf{DB}_{csat}}^{\mathbf{MQ}_{csat}}$ for which

$$\mathit{cnf}(\sigma(\mathbf{MQ}_{csat})) > \frac{k'-1}{2^h}$$

Obviously, the predicate variables $P_i$ $(1 \leq i \leq s)$ are mapped in $\sigma$ to either $p^a$ or $p^b$. In order to build an assignment $\rho$ which makes $\langle F, k', \Pi, \chi \rangle$ a YES instance, let $\rho(p_i) = \mathit{true}$ if, within $\sigma$, $P_i \mapsto p^a$ and $\rho(p_i) = \mathit{false}$ if $P_i \mapsto p^b$. Observe, then, that the join $J$ of body atoms of $\sigma(\mathbf{MQ}_{csat})$ has precisely $2^h$ tuples. To have $\mathit{cnf}(\sigma(\mathbf{MQ}_{csat})) > \frac{k'-1}{2^h}$, the join $J_h$ of body and head atoms of $\sigma(\mathbf{MQ}_{csat})$ must have at least $k'$ tuples, each of which represents one of the $k'$ assignments for the variables within $\chi$ that make $F$ true. This closes the proof. □

**Theorem 3.29** *The combined complexity of $\langle \mathbf{DB}, \mathbf{MQ}, \mathrm{cnf}, k, 1 \rangle$ and $\langle \mathbf{DB}, \mathbf{MQ}, \mathrm{cnf}, k, 2 \rangle$ is $NP^{PP}$ complete.*

**Proof.** The proof works using a slightly different reduction than the one used in the proof of the previous theorem, because here we have to deal with ordinary variables permutation/addition through instantiations. Suppose an instance of $\exists$**C-3SAT** is given, i.e.

- a formula $F = \bigwedge_{i=1}^{n} c_i$, where each $c_i$ is a three literal clause $l_{i_1} \vee l_{i_2} \vee l_{i_3}$ (with each $l_{i_j} \in \{p_1, \ldots, p_s, q_1, \ldots, q_h, \neg p_1, \ldots, \neg p_s, \neg q_1, \ldots, \neg q_h\}$,

- a partition of variables of $F$ in two sets $\Pi = \{p_1, \ldots p_s\}$ and $\chi = \{q_1, \ldots q_h\}$,

- an integer $k'$.

We build a database $\mathbf{DB}_{csat}$ and a metaquery $\mathbf{MQ}_{csat}$ as follows. $\mathbf{DB}_{csat}$ contains:

- a relation $p(X, \overline{X}, Y) = \{\langle 1, 0, l \rangle\}$;

- a relation $q(X, \overline{X}) = \{\langle 1, 0 \rangle, \langle 0, 1 \rangle\}$;

- a relation $ch(Y) = \{\langle l \rangle\}$;

- a relation

$$c'(L_1, L_2, L_3, C) = \begin{matrix} \{\langle 1,0,0,1\rangle, \\ \langle 0,1,0,1\rangle, \\ \langle 0,0,1,1\rangle, \\ \langle 1,0,1,1\rangle, \\ \langle 1,1,0,1\rangle, \\ \langle 0,1,1,1\rangle, \\ \langle 1,1,1,1\rangle, \\ \langle 0,0,0,0\rangle\} \end{matrix}$$

- the relation $c(C_1, \ldots, C_n) = \{\langle 1, \ldots, 1\rangle\}$.

As for $\mathbf{MQ}_{csat}$, we introduce for each boolean variable $p_i$ (resp. $q_i$) of $F$, two variables $P_i$ and $\overline{P}_i$ (resp. $Q_i$ and $\overline{Q}_i$), and $\mathbf{MQ}_{csat}$ is as follows:

$$c(C_1, \ldots, C_n) \leftarrow P'(P_1, \overline{P}_1, Y), \ldots, P'(P_s, \overline{P}_s, Y), ch(Y), q_1(Q_1, \overline{Q}_1), \ldots, q_h(Q_h, \overline{Q}_h),$$
$$c'(L_{1_1}, L_{1_2}, L_{1_3}, C_1), \ldots, c'(L_{n_1}, L_{n_2}, L_{n_3}, C_n).$$

where each $L_{i_j}$ ( $1 \leq i \leq n, 1 \leq j \leq 3$), will be

- either $P_y$ if $l_{i_j} = p_y$ and $p_y$ belongs to $\Pi$, or $\overline{P}_y$ if $l_{i_j} = \neg p_y$ and $p_y$ belongs to $\Pi$.

- either $Q_y$ if $l_{i_j} = q_y$ and $q_y$ belongs to $\chi$, or $\overline{Q}_y$ if $l_{i_j} = \neg q_y$ and $q_y$ belongs to $\chi$.

Finally, we set $k$ equal to $\frac{k'-1}{2^h}$.

Suppose that $\langle F, k', \Pi, \chi \rangle$ is a YES instance for $\exists$**C**-3SAT. Consider an assignment $\rho$ defined over variables $p_j$ ($1 \leq j \leq s$) which makes $\langle F, k', \Pi, \chi \rangle$ a YES instance. An instantiation $\sigma_{\mathbf{DB}_{csat}}^{\mathbf{MQ}_{csat}}$ (either type 1 or 2), which makes $\mathit{cnf}(\sigma(\mathbf{MQ}_{csat})) > \frac{k'-1}{2^h}$ can be built by mapping $P'$ to $p$, and mapping, for $1 \leq j \leq k$, $P_j$ to the first attribute of $p$, if $\rho(p_j) = \mathit{true}$, or, otherwise, to its second attribute. The join $J$ of body atoms of $\sigma(\mathbf{MQ}_{csat})$ will have precisely $2^h$ tuples. Moreover, if there are at least $k'$ assignments for variables within $\chi$, joining the head relations with $J$ will result in a relation $J_h$ (which selects only those tuples of $J$ with each $C_i$ variable set to 1, i.e. the clause $c_i$ is true) containing at least $k'$ tuples. Hence $\mathit{cnf}(\sigma(\mathbf{MQ}_{csat})) > k$.

Now, suppose that we have a type-1 (or type 2) instantiation $\sigma$ for $\mathbf{MQ}_{csat}$ for which

$$\mathit{cnf}(\sigma(\mathbf{MQ}_{csat})) > \frac{k'-1}{2^h}$$

(observe that the considerations that follow does not hold for type-0 instantiations). Note that the predicate variable $P'$ is necessarily mapped to $p$, indeed the only alternative instantiations for $P'$ associate it to $c$ or to $c'$; but, in this case, the empty relation is produced when an atom like, say, $c'(P_i, \overline{P}_i, Y)$ (produced from a relation pattern $P(P_i, \overline{P}_i, Y)$) is joined to $ch(Y)$.

In order to build an assignment $\rho$ which makes $\langle F, k', \Pi, \chi \rangle$ a YES instance, let $\rho(p_i) = \mathit{true}$, if $P_i$ is mapped to the first attribute of $p$, and $\rho(p_i) = \mathit{false}$ if $P_i$ is mapped to the second attribute of $p$. Observe that it cannot be the case that $P_i$ is mapped to the third attribute of $p$, because, otherwise, the resulting natural join with $ch(Y)$ (e.g. $p(\overline{P}_g, Y, P_g) \bowtie ch(Y)$), would be empty. Observe then, that the join $J$ of body atoms of $\sigma(\mathbf{MQ}_{csat})$ has precisely $2^h$ tuples. To have $cnf(\sigma(\mathbf{MQ}_{csat})) > \frac{k'-1}{2^h}$, the join $J_h$ of body and head atoms must have at least $k'$ tuples, each of which represents one of the $k'$ assignments for the variables of $\chi$ that make $F$ true. This completes the proof. $\square$

## 3.4 The complexity of acyclic metaqueries

We have shown above that, as far as the combined complexity measure is concerned, metaquerying is intractable. Next we discuss some tractable subcases. We recall some definitions first.

**Definition 3.30** A *Hypergraph* $H = \langle V, E \rangle$ is a set $V$ of vertices, and a set $E \subseteq 2^V$ of (hyper)edges. An *ear* for a hypergraph $H = \langle V, E \rangle$ is an edge $e \in E$ such that for some distinct edge $w \in E$, called the *witness* of $e$, no vertex of $e - w$ is in any other edge. We say that $H$ is acyclic if its derived hypergraph $GYO(H)$ is empty, where $GYO(H)$ is built by applying the following steps until there are no ears (see [1]) :

1. remove from $E$ all isolated edges, i.e., edges sharing no vertex with other edges.

2. choose an ear $e$ of $H$.

3. remove $e$ from $H$ by deleting it from $E$ and by deleting from $V$ vertices of $e$ not appearing elsewhere in $E$.

The following notion of acyclicity for metaqueries is a little bit different from the usual one employed for ordinary conjunctive queries [7]. In particular, the hypergraph associated with a metaquery will contain both ordinary and predicate variables as nodes.

**Definition 3.31** Let $s$ be an atom, a relation pattern, or a metaquery. The set of (both predicate and ordinary) variables of $s$ is denoted $var(s)$, whereas the set of ordinary variables of $s$ is denoted $var_o(s)$. Let $\mathbf{MQ}$ be a metaquery, define the *hypergraph* $H(\mathbf{MQ}) = \langle V, E \rangle$ associated with $\mathbf{MQ}$ as follows. $V = var(\mathbf{MQ})$, whereas $E$ contains an edge $e_i = var(R_i(\mathbf{X}_i))$ for each literal scheme $R_i(\mathbf{X}_i)$ in $\mathbf{MQ}$. We say that $\mathbf{MQ}$ is *acyclic* if $H(\mathbf{MQ})$ is acyclic. Similarly, define the *semi-hypergraph* $SH(\mathbf{MQ}) = \langle V', E' \rangle$ associated with $\mathbf{MQ}$ as follows. $V' = var_o(\mathbf{MQ}))$, $E'$ contains an edge $e_i = var_o(R_i(\mathbf{X}_i)$ for each literal scheme $R_i(\mathbf{X}_i)$ in $\mathbf{MQ}$. We say that $\mathbf{MQ}$ is *semi-acyclic* if $SH(\mathbf{MQ})$ is acyclic.

For example, the metaquery

$$\mathbf{MQ}_1 = P(X,Y) \leftarrow P(Y,Z), Q(Z,W)$$

is acyclic, whereas the slightly different metaquery

$$\mathbf{MQ}_2 = P(X,Y) \leftarrow Q(Y,Z), P(Z,W)$$

is cyclic. Finally, the metaquery

$$\mathbf{MQ}_1 = N(X) \leftarrow N(Y), E(X, Y)$$

is semi-acyclic, but it is not acyclic. It is straightforward to show that an acyclic metaquery is semi-acyclic as well.

**Theorem 3.32** *Let* **MQ** *be an acyclic metaquery and* $I \in \mathbf{I}$ *a plausibility index. The combined complexity of* $\langle \mathbf{DB}, \mathbf{MQ}, I, 0, 0 \rangle$ *is LOGCFL-complete under logspace reductions.*

**Proof.** *Hardness.* The satisfiability problem for acyclic conjunctive queries was proved hard for *LOGCFL* in [18]. Thus, let $\mathbf{Q} = Q_1, \ldots, Q_n$ be an acyclic conjunctive query, where each $Q_i$ is an atom. The (completely instantiated) metaquery

$$mq(\mathbf{Q}) = Q_1 \leftarrow Q_1, Q_2, \ldots, Q_n \tag{5}$$

is acyclic. By Proposition 3.20, the form of $mq(\mathbf{Q})$ is such that the BCQ problem on $\mathbf{Q}$ can be solved by showing a positive value of either cover, support or confidence for $mq(\mathbf{Q})$.

*Membership.* **MQ** can be easily reduced, in logspace, to an acyclic conjunctive query $\mathbf{Q_{MQ}}$, defined on a new database $D_{\mathbf{DB}}$. Let $r$ be a relation, we denote by $a(r)$ the corresponding arity. The reduction is as follows:

- For each relation name $r$ in **DB** we introduce a new constant value $n_r$;

- For each arity $a$ of some relation of **DB**, we introduce a relation $u_a$, of arity $a+1$. The extension of $u_a$ in $D_{\mathbf{DB}}$ is such that if $t = \langle t_1, ..., t_a \rangle$ is a tuple belonging to a relation $r$ of arity $a$ then the tuple $\langle n_r, t_1, ..., t_a \rangle$ is in $u_a$.

- Assume that **MQ** is of the form

  $T(T_1, .., T_{a(T)}) \leftarrow L_1(X_1^1, .., X_{a(L_1)}^1), .., L_m(X_1^m, .., X_{a(L_m)}^m)$

  then if $I \neq sup$, we set $\mathbf{Q_{MQ}}$ to

  $u_{a(T)}(T, T_1, .., T_{a(T)}), u_{a(L_1)}(L_1, X_1^1, .., X_{a(L_1)}^1), ..., u_{a(L_m)}(L_m, X_1^m, .., X_{a(L_m)}^m).$

  If $I = sup$, we set $\mathbf{Q_{MQ}}$ to

  $u_{a(L_1)}(L_1, X_1^1, .., X_{a(L_1)}^1), ..., u_{a(L_m)}(L_m, X_1^m, .., X_{a(L_m)}^m).$

It is easy to see that an instance of $\langle \mathbf{DB}, \mathbf{MQ}, I, 0, 0 \rangle$ evaluates to *true* iff $\mathbf{Q_{MQ}}$ has a non-empty answer over $D_{\mathbf{DB}}$. □

However, acyclicity is not sufficient to guarantee tractability in general, as shown next for instantiation types other than type-0.

**Theorem 3.33** *Let $I \in \mathbf{I}$ be a plausibility index and $T \in \{1, 2\}$ and let $\mathbf{MQ}$ be an acyclic metaquery. The combined complexity of $\langle \mathbf{DB}, \mathbf{MQ}, I, 0, T \rangle$ is NP-complete.*

**Proof.** Membership follows from Theorem 3.21. As for hardness, we show that we can reduce the *NP*-complete problem HAMILTONIAN PATH [16] to $\langle \mathbf{DB}, \mathbf{MQ}, I, 0, T \rangle$. An instance of HAMILTONIAN PATH deals with an undirected graph $G = \langle V, E \rangle$, and asks if $G$ contains a Hamiltonian path, i.e. a path touching each node in V exactly once. Without losing generality, we can assume that $|V| > 2$.

We build a database $\mathbf{DB}_{ham}$ and a metaquery $\mathbf{MQ}_{ham}$. $\mathbf{DB}_{ham}$ contains a relation $g$, with a single tuple encoding node names, say $t = \langle v_1, \ldots, v_n \rangle$, and a binary relation $e$, storing one tuple for each edge in $E$. The metaquery $\mathbf{MQ}_{ham}$ is

$$N(X_1, .., X_n) \leftarrow N(X_1, .., X_n), e(X_1, X_2), .., e(X_{n-1}, X_n)$$

Intuitively, we use $N$ to select a permutation of nodes of $G$ and the body of $\mathbf{MQ}$ encodes the constructed Hamiltonian path. Now, suppose an Hamiltonian path $p = \langle v_{u_1}, \ldots, v_{u_n} \rangle$ exists: we can build from $p$ the following set of ground atoms:

$$\{g(v_{u_1}, \ldots, v_{u_n}), e(v_{u_1}, v_{u_2}), \ldots, e(v_{u_{n-1}}, v_{u_n})\}$$

which is satisfied over $\mathbf{DB}_{ham}$. By Proposition 3.20, this certifies that $\langle \mathbf{DB}_{ham}, \mathbf{MQ}_{ham}, I, 0, T \rangle$ is a YES instance, for any $I \in \mathbf{I}$. Similarly, observing that $N$ can suitably match only with $g$ (both for type-1 and for type-2 instantiations), a YES-certificate for $\langle \mathbf{DB}_{ham}, \mathbf{MQ}_{ham}, I, 0, T \rangle$ allows us to build an Hamiltonian path of $G$. Moreover, $\mathbf{MQ}$ is acyclic: take $\{N, X_1, \ldots, X_n\}$ as witness for $\{X_i, X_{i+1}\}$, $1 \leq i \leq n-1$.  □

**Theorem 3.34** *Let $I \in \mathbf{I}$ be a plausibility index and $T \in \{1, 2\}$ and let $\mathbf{MQ}$ be an acyclic metaquery. The combined complexity of $\langle \mathbf{DB}, \mathbf{MQ}, \sup, k, T \rangle$ and $\langle \mathbf{DB}, \mathbf{MQ}, \text{cvr}, k, T \rangle$ is NP-complete.*

**Proof.** Membership follows from Theorem 3.24, whereas hardness is by reduction from $\langle \mathbf{DB}, \mathbf{MQ}, sup, 0, T \rangle$ to $\langle \mathbf{DB}, \mathbf{MQ}, sup, k, T \rangle$ and from $\langle \mathbf{DB}, \mathbf{MQ}, cvr, 0, T \rangle$ to $\langle \mathbf{DB}, \mathbf{MQ}, cvr, k, T \rangle$, respectively.  □

One can ask if disregarding predicate variables, and thus referring to semi-acyclic metaqueries, instead of considering acyclic ones would be sufficient in order to give a polynomial evaluation algorithm as far as type-0 metaqueries are concerned. The next result shows that, unfortunately, the evaluation of semi-acyclic type-0 metaqueries is not simpler than evaluating general metaqueries.

**Theorem 3.35** *Let $I \in \mathbf{I}$ be a plausibility index and let $\mathbf{MQ}$ be a semi-acyclic metaquery. The combined complexity of $\langle \mathbf{DB}, \mathbf{MQ}, I, 0, 0 \rangle$ is NP-complete.*

**Proof.** Membership follows from Theorem 3.21. As for hardness, we proceed by reducing the *NP*-complete problem 3-COLORING [16] to the problem of evaluating type-0 semi-acyclic metaqueries. Let $G = \langle V, E \rangle$ be an undirected graph. We build a database $\mathbf{DB}_{3col}$ and a metaquery $\mathbf{MQ}_{3col}$ as follows. $\mathbf{DB}_{3col}$ contains three binary relations $r', g'$ and $b'$, with the following extensions: $r'(X,Y) = \{\langle g,r \rangle, \langle b,r \rangle\}, g'(X,Y) = \{\langle r,g \rangle, \langle b,g \rangle\}, b'(X,Y) = \{\langle g,b \rangle, \langle r,b \rangle\}$.

In $\mathbf{MQ}_{3col}$ we use ordinary variables $X_u$ and predicate variables $X'_u$, one for each node $u \in V$. We denote by "_" a new variable not occurring elsewhere in $\mathbf{MQ}_{3col}$ (a *mute* variable).

Let $E = \{(u_1, v_1), \ldots, (u_m, v_m)\}$, and $V = \{z_1, \ldots, z_n\}$. Then $\mathbf{MQ}_{3col}$ contains both the following sets of literals

$$S' \equiv \{X'_{u_1}(X_{v_1}, \_), \ldots, X'_{u_m}(X_{v_m}, \_)\}$$

$$S'' \equiv \{X'_{z_1}(\_, X_{z_1}), \ldots, X'_{z_n}(\_, X_{z_n})\}$$

Intuitively, $S'$ has the same purpose as in Theorem 3.21, that is, it encodes $G$ as a set of literals. The atoms of $S''$ will force each variable $X'_{z_i}$ to represents the same color of the corresponding $X_{z_i}$ variable. $\mathbf{MQ}_{3col}$ is as follows

$$X'_{u_1}(X_{v_1}, \_) \leftarrow X'_{u_1}(X_{v_1}, \_), \ldots, X'_{u_m}(X_{v_m}, \_), X'_{z_1}(\_, X_{z_1}), \ldots, X'_{z_n}(\_, X_{z_n})$$

As an example, let $G = \langle \{1, 2, 3\}, \{(1, 2), (3, 1), (3, 2)\}\rangle$. In this case, $\mathbf{MQ}_{3col}$ is

$$X'_1(X_2, \_) \leftarrow X'_1(X_2, \_), X'_3(X_1, \_), X'_3(X_2, \_), X'_1(\_, X_1), X'_2(\_, X_2), X'_3(\_, X_3)$$

Note that, in general, $\mathbf{MQ}_{3col}$ might not be acyclic, but it is semi-acyclic. In fact, let $SH(\mathbf{MQ}_{3col}) = \langle H_v, H_e \rangle$ be the semi-hypergraph associated to $\mathbf{MQ}_{3col}$. $H_v$ contains:

- the set $\Lambda$ of variables $X_{z_1}, \ldots, X_{z_n}$;
- a set $\Phi$ of mute variables $\phi_1, \ldots, \phi_m$, one for each atom of $S'$;
- a set $\Psi$ of mute variables $\psi_1, \ldots, \psi_n$, one for each atom of $S''$.

$H_e$ is a set of hyperedges either of the form $(X_i, \phi_j)$, or of the form $(\psi_i, X_i)$. Since variables of $\Psi$ and variables of $\Phi$ appear in at most one hyperedge, and there is no hyperedge sharing two different variables of $\Lambda$, we can conclude that $SH(\mathbf{MQ}_{3col})$ is acyclic, and hence $\mathbf{MQ}_{3col}$ is semi-acyclic.

Similar to the proof of Theorem 3.21, we conclude this proof by showing the following:

**Claim 3.36** *$G$ has a 3-coloring iff there exists a type-0 instantiation $\sigma_{\mathbf{DB}_{3col}}^{\mathbf{MQ}_{3col}}$ such that $\sigma(S')$ has a ground instance $s'$ and $\sigma(S'')$ has a ground instance $s''$ such that $s' \cup s''$ is true in $\mathbf{DB}$.*

($\rightarrow$). Suppose $G$ has a 3-coloring $c : V \mapsto \{r, g, b\}$, defined for each $u \in V$, and such that if $(u, v) \in E$ then $c(u) \neq c(v)$. Let $c' : V \mapsto \{r', g', b'\}$ be a coloring isomorphic to $c$. Consider then an instantiation $\sigma$ which maps each literal in the form $X'_u(X_v, \phi)$ to an atom in the form $c'(u)(X_v, \phi)$, and each atom in the form $X'_z(\psi, X_z)$ to an atom in the form $c'(z)(\psi, X_z)$, where $u' \in \{r', g', b'\}$. A ground instance $s' \cup s''$ of variables for $\sigma(S' \cup S'')$ which is true in **DB** is

$$\underbrace{c'(u_1)(c(v_1), c(u_1)), \ldots, c'(u_m)(c(v_m), c(u_m))}_{s'}, \underbrace{c'(z_1)(\beta_1, c(z_1)), \ldots, c'(z_n)(\beta_n, c(z_n))}_{s''}$$

where constants $\beta_1, \ldots, \beta_n$ are chosen accordingly.

($\leftarrow$). Suppose that there is an instantiation $\sigma$ of type 0 for $\mathbf{MQ}_{3col}$ that makes a given ground instance $s' \cup s''$ of $\sigma(S' \cup S'')$ true in $\mathbf{DB}_{3col}$. $\sigma$ maps each predicate variable $X'_u$ to either $r'$, $g'$ or $b'$. $s' \cup s''$ has the form

$$s' \cup s'' \equiv \underbrace{c'_{u_1}(c_{v_1}, \alpha_1), \ldots, c'_{u_m}(c_{v_m}, \alpha_m)}_{s'}, \underbrace{c'_{z_1}(\beta_1, c_{z_1}), \ldots, c'_{z_n}(\beta_n, c_{z_n})}_{s''}$$

where each constant value $c_{z_i}, c_{v_j}$ $(1 \leq j \leq m), (1 \leq i \leq n)$ belongs to $\{r, g, b\}$, and each constant $c'_{u_j}$ or $c'_{z_i}$, $(1 \leq j \leq m)$ $(1 \leq i \leq n)$, belongs to $\{r', g', b'\}$. Consider now the following mapping:

$$M_{S' \cup S'', s' \cup s''} \equiv \{X'_{u_1}(X_{v_1}, \phi_1) \mapsto c'_{u_1}(c_{v_1}, c_{u_1}), \ldots, X'_{u_m}(X_{v_m}, \phi_m) \mapsto c'_{u_m}(c_{v_m}, c_{u_m}),$$
$$X'_{z_1}(\psi_1, X_{z_1}) \mapsto c'_{z_1}(\beta_1, c_{z_1}), \ldots, X'_{z_n}(\psi_n, X_{z_n}) \mapsto c'_{z_n}(\beta_n, c_{z_n})\}$$

from $S$ to $s' \cup s''$. Note that if a constant $c_u$ is $r$ then $c'_u$ is $r'$, whereas if a constant $c_u$ is $g$ (resp. $b$) then $c'_u$ is $g'$ (resp. $b'$). Since it cannot be the case that $c_{v_1} \neq c_{u_1}$ for an atom $c'_{u_1}(c_{v_1}, c_{u_1}) \in s'$, we can build a valid 3-coloring $c : V \mapsto \{1, 2, 3\}$ by setting $c(z_i) = c_{z_i}$ for each $z_i \in V$. This closes the proof of the Claim. We can then resume the proof of Theorem 3.35.

The Theorem follows by noting that $MQ_{3col}$ is built in order for $\sigma(S' \cup S'')$ to be a certifying set for $cvr, sup$ and $cnf$, for any instantiation $\sigma_{\mathbf{DB}_{3col}}^{\mathbf{MQ}_{3col}}$, and, therefore, $cvr(\sigma) > 0$, $sup(\sigma) > 0$ and $cnf(\sigma) > 0$ iff claim 3.36 holds. □

## 3.5 Data complexity

In this section we discuss data complexity of metaquerying problems. Similarly to most query languages, the data complexity is much lower than the combined complexity. In particular, in some interesting case, it lies very low in the complexity hierarchy, as proven next.

**Theorem 3.37** *Under the data complexity measure (fixed metaquery and threshold value, variable database), $\langle \mathbf{DB}, \mathbf{MQ}, I, 0, T \rangle$ is in $AC^0$, for $I \in \mathbf{I}$ and for $T \in \{0, 1, 2\}$.*

**Proof.** Under the data complexity measure, number of type-T instantiations for $\mathbf{MQ}$ and $\mathbf{DB}$ is constant.

We can build a circuit solving the instance $\langle \mathbf{DB}, \mathbf{MQ}, I, 0, T \rangle$ at hand as follows: Let $\sigma_{\mathbf{DB}}^{\mathbf{MQ}}$ be a generic instantiation of $\mathbf{MQ}$; let $\sigma(\mathbf{MQ}) = Q_1 \leftarrow Q_2, \ldots, Q_n$, where the $Q_i$'s are atoms; let $q'(\sigma(\mathbf{MQ}))$ and $q''(\sigma(\mathbf{MQ}))$ denote the boolean conjunctive queries $\mathbf{Q}' = Q_1, \ldots, Q_n$ and $\mathbf{Q}'' = Q_2, \ldots, Q_n$, respectively. By Proposition 3.20, for $I \in \{cnf, cvr\}$ (resp. $I = sup$), $I(\sigma(\mathbf{MQ})) > 0$ iff $\mathbf{Q}'$ (resp. $\mathbf{Q}''$) is satisfiable in $\mathbf{DB}$. Now, it is known (see [6]) that any conjunctive query $\mathbf{Q}$ is solved by a constant-depth polynomial size logspace-uniform family of boolean circuits, call it $\{C(\mathbf{Q})_i\}$, where, for each $i$, $C(\mathbf{Q})_i$ solves $\mathbf{Q}$ when the input database instance has size $i$. Let $\Sigma^T$ be the set of all type-T instantiations for the metaquery at hand. For a fixed database size $i$, consider the circuit $C(\mathbf{MQ})_i$ obtained by connecting the outputs of all the circuits $C(q'(\sigma(\mathbf{MQ}))_i$ (resp. $C(q''(\sigma(\mathbf{MQ}))_i)$, with $\sigma \in \Sigma^T$ through an OR gate. Since $|\Sigma^T|$ is polynomial, $C(\mathbf{MQ})_i$ has constant depth and polynomial size. Hence the result follows. □

In the general case, the data complexity of metaquerying is within $TC^0$.

**Theorem 3.38** *Let $T \in \{0, 1, 2\}$. Let $\mathbf{MQ}$ be a metaquery, $\mathbf{DB}$ be a database, $I \in \mathbf{I}$ a plausibility index, and $0 \leq k < 1$. The data complexity of the metaquerying problem $\langle \mathbf{DB}, \mathbf{MQ}, I, k, T \rangle$ is in $TC^0$.*

**Proof.** Any project-join expression $\mathbf{Q}$ is solved by a constant-depth polynomial-size logspace-uniform family of multiple-output circuits of unbounded fan-in AND, OR, and NOT gates (see [1]); call the generic circuit of this family $\{C'(\mathbf{Q})_i\}$, where, for each $i$, $C'(\mathbf{Q})_i$ calculates the output of $\mathbf{Q}$ when the input database instance has size $i$. In particular each $C'(\mathbf{Q})_i$ has $M_i$ boolean outputs, one for each tuple potentially in the result of $\mathbf{Q}$. Note that, under data the complexity measure, for each database size $i$, $M_i$ is polynomial in $i$.

Let $C'_j(\mathbf{Q})_i$ be the subcircuit of $C'(\mathbf{Q})_i$ deciding if the $j$th tuple in the potential result of $\mathbf{Q}$ belongs to its output, for $j = 1, \ldots, M_i$. Since every language in $AC^0$ has its characteristic function in $\#AC^0$ [2], there exists a $\#AC^0$ circuit $C''_j(\mathbf{Q})_i$ equivalent to $C'_j(\mathbf{Q})_i$, for each $j = 1, \ldots, M_i$.

Let $\{count(\mathbf{Q})_i\}$ be the family of circuits obtained connecting the outputs of the circuits $C''_j(\mathbf{Q})_i$, for each $j = 1, \ldots, M_i$, to a single +-gate, for each $i \geq 0$. Then $\{count(\mathbf{Q})_i\}$ is a family of $\#AC^0$ circuits computing $|\mathbf{Q}|$. Next, we prove a technical lemma, that we will use to conclude the proof.

**Lemma 3.39** *Let $r$ be a Horn rule, $\mathbf{Q_n}$ and $\mathbf{Q_d}$ be two, polynomial sized, project-join expressions defined over $\mathbf{DB}$ and the predicates of $r$, and $k$ be a rational value (such that $0 \leq k < 1$, and $k$ encoded as a pair of naturals $(a,b)$, such that $k = \frac{a}{b}$). Then there exists a constant-depth polynomial-size (w.r.t. the size of $\mathbf{DB}$) uniform family $\{C(r)_i\}$ of circuits of unbounded fan-in NOT and MAJORITY gates, such that $C(r)_i$ outputs 1, iff $\frac{|\mathbf{Q_n}|}{|\mathbf{Q_d}|} > k$, where $|\mathbf{DB}| = i$.*

**Proof.** Consider the function $f(\mathbf{DB}, r) = b|\mathbf{Q_n}| - a|\mathbf{Q_d}|$ taking value in $\mathbf{N}$. Clearly, $\frac{|\mathbf{Q_n}|}{|\mathbf{Q_d}|} > k$ iff $f(\mathbf{DB}, r) > 0$. We recall the following result of [4]: for each integer $N$ there exists a log-time uniform $\#AC^0$ circuit, which, having in input the binary representation of $N$, computes $N$. Call this circuit $number(N)$. Since $a$ and $b$ are integers, it is easy to build two $\#AC^0$ circuits computing the functions $b|\mathbf{Q_n}|$ and $a|\mathbf{Q_d}|$, connecting $number(b)$ to $count(\mathbf{Q_n})_i$ (resp. $number(a)$ to $count(\mathbf{Q_d})_i$) through a $\times$-gate. Then, the function $f$ is in the class $GapAC^0$, and the language $\{r \mid \frac{|Q_n|}{|Q_d|} > k \text{ over } \mathbf{DB}\}$, where $r$ is a Horn rule over $\mathbf{DB}$, is in the class $PAC^0 = TC^0$.

Therefore there exists a constant-depth polynomial-size uniform family $\{C(r)_i\}$ of circuits of unbounded fan-in MAJORITY and NOT gates, such that $C(r)_i$ outputs 1 iff $\frac{|Q_n|}{|Q_d|} > k$, when the input database instance has size $i$. □

In order to conclude our proof, we build a $TC^0$ circuit family $\{C(r)_i\}$, for $I = cvr$, $I = cnf$ and $I = sup$, respectively, as follows:

- For $I = cvr$ and $I = cnf$, since both these indices have the form $\frac{|\mathbf{Q_n}|}{|\mathbf{Q_d}|}$, $\{C(r)_i\}$ is given as in Lemma 3.39;

- for $I = sup$, let $r$ be in the form $t(\mathbf{X}_0) \leftarrow l_1(\mathbf{X}_1), \ldots, l_n(\mathbf{X}_n)$. Consider the set of circuits families $\mathcal{C} = \{\{C^1(r)_i\}, \ldots, \{C^n(r)_i\}\}$, where each family $\{C^j(r)_i\}$ ($1 \leq j \leq n$), computes the function $f_j(\mathbf{DB}, r) = a|\mathbf{Q_j}| - b|\mathbf{Q'_j}|$, where $\mathbf{Q_j}$ is the natural join of $l_1, \ldots, l_n$, projected over the attributes of $l_j$, and $\mathbf{Q'_j}$ is simply $l_j$. By Lemma 3.39, each $C^j(r)_i \in \mathcal{C}$ ( $1 \leq j \leq n$ ) is $TC^0$ and outputs 1 iff the corresponding ratio $\frac{|\mathbf{Q_j}|}{|\mathbf{Q_d}|}$ is greater than $k$, when the input database instance has size $i$. Connecting the $n$ (a constant value) circuits for each $i$, through an OR gate, we obtain the needed circuit family $\{C(r)_i\}$.

Hence the result follows. □

**Remark**. Note that the proofs of Theorem 3.37 and Theorem 3.38 outline a general strategy for proving membership in $AC^0/TC^0$ for those indices either defined or reducible to a ratio between two project-join expressions defined (and polynomially sized) over the atoms of a given instantiated metaquery.

## 4 Algorithms for metaquery answering

In this section we address the problem of efficiently computing all the instantiations $\sigma_{\mathbf{DB}}^{\mathbf{MQ}}$ of a fixed metaquery $\mathbf{MQ}$ on an input database $\mathbf{DB}$, such that the values $cnf(\sigma(\mathbf{MQ}))$, $cvr(\sigma(\mathbf{MQ}))$, and $sup(\sigma(\mathbf{MQ}))$ are greater than user-provided thresholds, $k_{cnf}$, $k_{cvr}$, and $k_{sup}$, resp.

We decompose our answering problem into the following three subproblems:

1. Find all the partial instantiations $\sigma_b$ defined on $body(\mathbf{MQ})$ such that the support obtained by applying $\sigma_b$ to $\mathbf{MQ}$ is greater than $k_{sup}$ in $\mathbf{DB}$.

2. For each partial instantiation $\sigma_b$ found in step 1, find all the partial instantiations $\sigma_h$ defined on $head(\mathbf{MQ})$ such that $\sigma = \sigma_h \circ \sigma_b$ is an instantiation defined on $\mathbf{MQ}$, and $cvr(\sigma(\mathbf{MQ})) > k_{cvr}$.

3. Return as a solution the instantiations $\sigma$ found in step 2 such that $cnf(\sigma(\mathbf{MQ})) > k_{cnf}$.

The rationale behind this choice is as follows. First, we note that a high-support body is potentially shared by various rules, having different heads. Furthermore, by definition, the computation of the support requires to reduce, one after another, the relations in the rule's body, that is, to compute $\pi_{att(r_i)}(\mathbf{J}(body(r)))$, for each $r_i \in body(r)$; reducing a relation $r_i$ w.r.t. a set of relations $S$ can be sometimes (i.e., when some acyclicity criterion is met) performed without explicitly computing $\mathbf{J}(S)$, hereby gaining in efficiency. Similarly, the computation of the cover (step 2) requires only the reduction of the relation associated to the head. Finally, the computation of the overall join (step 3) becomes less expensive if the involved relations are reduced (see [28]).

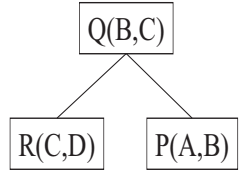

Figure 3: The join tree of Example 4.3.

Next, we recall and/or extend some definitions related to computing conjunctive queries. For simplicity of notation, as we have also done above, in the following we refer to atoms and to associated relations interchangeably, unless ambiguity arises.

**Definition 4.1** Let $r_1, \ldots, r_n$ be a set of relations. We say that $r_i$, $i \in \{1, ..., n\}$, is *reduced* w.r.t. $r_1, \ldots, r_n$, if $r_i = \pi_{att(r_i)}(r_1 \bowtie \ldots \bowtie r_n)$. We say that $r_1, \ldots, r_n$ is *reduced* if $r_i$ is reduced w.r.t. $r_1, \ldots, r_n$, for each $i = 1, \ldots, n$.

**Definition 4.2** A *join tree* for a set of literal schemes $Q$ is a tree $\mathcal{T}$, whose vertices are the literal schemes of $Q$, such that whenever the same ordinary variable $X$ occurs in two literal schemes $L_1$ and $L_2$, then $X$ occurs in each literal scheme on the unique path linking $L_1$ and $L_2$ (see [1]).

**Example 4.3** Let $Q$ be the set $\{P(A,B), Q(B,C), R(C,D)\}$ of literal schemes. Then a join tree for $Q$ is reported in Figure 3.

It is then easy to generalize analogous results proved for conjunctive queries, to show that a metaquery is semi-acyclic iff its set of literals has a join tree [1, 18].

**Definition 4.4** Let $Q = \{r_1, \ldots, r_n\}$ be a set of atoms on a database **DB**. A *semijoin step* is an expression of the form $r_i := r_i \ltimes r_j$, with $1 \leq i, j \leq n$. A *semijoin program* is a sequence of semijoin steps. A semijoin program is called a *full reducer* for $Q$ if, after executing that program, each $r_i$ is reduced, independently of the initial values of the relations (see [1]).

A full reducer is a way to efficiently reduce a set of atoms. Unfortunately, not every set of atoms has a full reducer. Indeed, it is proved in [9] that a set of atoms has a full reducer iff it is semi-acyclic.

For a semi-acyclic set of atoms $Q$, a full reducer consists of a sequence of two semijoin programs of the same length, called *first-half* and *second-half* respectively. Let $\mathcal{T}$ be a rooted join tree for $Q$. The first-half is obtained by performing a bottom-up visit of $\mathcal{T}$: let $r_i$ be the current node in the visit, then for each child $r_j$ of $r_i$ in $\mathcal{T}$, add $r_i := r_i \ltimes r_j$ as next step of the sequence. The second-half is then obtained from the first-half, by reversing the sequence and exchanging the relations, i.e. from $r_i := r_i \ltimes r_j$ we obtain $r_j := r_j \ltimes r_i$.

**Example 4.5** Consider the set $Q = \{p(A,B), p(B,C), r(C,D)\}$. Let $\mathcal{T} = \langle Q, \{\langle q(B,C), p(A,B)\rangle, \langle q(B,C), r(C,D)\rangle\}\rangle$ be an associated join tree. Suppose $\mathcal{T}$ is rooted at $q(B,C)$. A full reducer for $Q$ is:

$$\left.\begin{array}{l} q(B,C) := q(B,C) \ltimes r(C,D) \\ q(B,C) := q(B,C) \ltimes p(A,B) \end{array}\right\} \text{first-half}$$

$$\left.\begin{array}{l} p(A,B) := p(A,B) \ltimes q(B,C) \\ r(C,D) := r(C,D) \ltimes q(B,C) \end{array}\right\} \text{second-half}$$

In general, to compute the value of support, we need to reduce a generic set of atoms, not necessarily a semi-acyclic one. To achieve this goal efficiently we exploit the concept of *hypertree decomposition* of a conjunctive query [17] (see also [18] for a detailed discussion on the concept of "degree of cyclicity" of a conjunctive query).

**Definition 4.6** Let $Q$ be a set of literal schemes. A *hypertree* for $Q$ is a triple $\langle \mathcal{T}, \chi, \lambda \rangle$, where $\mathcal{T}$ is a rooted tree, and $\chi$ and $\lambda$ are labeling functions which associate to each vertex $p$ of $\mathcal{T}$ two sets $\chi(p) \subseteq var_o(Q)$ and $\lambda(p) \subseteq Q$.

**Definition 4.7** Let $\mathcal{T}$ be a rooted tree. We denote by $vertices(\mathcal{T})$ the set of nodes of $\mathcal{T}$. Define $\chi(\mathcal{T})$ as $\cup_{p \in vertices(\mathcal{T})} \chi(p)$. For any $p \in vertices(\mathcal{T})$, $\mathcal{T}_p$ denotes the subtree of $\mathcal{T}$ rooted at $p$. A *hypertree decomposition* [17] of a set of literal schemes $Q$ is an hypertree for $Q$ such that

1. for each literal scheme $L \in Q$, there exists $p \in vertices(\mathcal{T})$ such that $var_o(L) \subseteq \chi(p)$;

2. for each ordinary variable $Y \in var_o(Q)$, the set $\{p \in vertices(\mathcal{T}) \mid Y \in \chi(p)\}$ induces a (connected) subtree of $\mathcal{T}$;

3. for each vertex $p \in vertices(\mathcal{T})$, $\chi(p) \subseteq var_o(\lambda(p))$;

4. for each vertex $p \in vertices(\mathcal{T})$, $var_o(\lambda(p)) \cap \chi(\mathcal{T}_p) \subseteq \chi(p)$.

An hypertree decomposition $\langle \mathcal{T}, \chi, \lambda \rangle$ of $Q$ is *complete* if, for each $L \in Q$, there exists $p \in vertices(\mathcal{T})$ such that $var_o(L) \subseteq \chi(p)$ and $L \in \lambda(p)$.

We refer to [17] for a complete discussion about the hypertree decomposition of conjunctive queries.

**Example 4.8** Let $Q^{ex}$ be the set $\{P(A,B), Q(B,C), R(C,D), S(B,D)\}$ of literal schemes. Let $\mathcal{T}^{ex}$ be the hypertree $\langle \{\langle p_1, p_2\rangle, \langle p_2, p_3\rangle\}, \chi^{ex}, \lambda^{ex} \rangle$ rooted at $p_1$, where $\chi^{ex}(p_1) = \{A,B\}$, $\chi^{ex}(p_2) = \{B,C\}$, $\chi^{ex}(p_3) = \{B,C,D\}$, $\lambda^{ex}(p_1) = \{P(A,B)\}$, $\lambda^{ex}(p_2) = \{Q(B,C)\}$, and $\lambda^{ex}(p_3) = \{R(C,D), S(B,D)\}$. Then $\langle \mathcal{T}^{ex}, \chi^{ex}, \lambda^{ex}\rangle$ is a hypertree decomposition of $Q^{ex}$.

The *width* of an hypertree decomposition $\langle \mathcal{T}, \chi, \lambda \rangle$ is defined as $\max_{p \in vertices(\mathcal{T})} |\lambda(p)|$. The *hypertree-width* $hw(Q)$ of a set of literal schemes $Q$ is defined as the minimum width over all its hypertree decompositions.

The notion of bounded hypertree-width generalizes the notion of semi-acyclicity for metaqueries. Indeed, a set of literal schemes $Q$ is semi-acyclic iff $hw(Q) = 1$.

```
function FINDRULES(DB, MQ, k_cnf, k_cvr, k_sup, T) : set of instantiations;
var    Σ : set of instantiations;
       r, s : array [1..n] of relation;
       procedure FINDHEADS(σ_b : an instantiation);
           function ENOUGHSUPPORT : boolean;
           begin
               for each literal scheme a ∈ body(MQ) do begin
                   let i such that var_o(a) ⊆ χ(p_{ν(i)}) and a ∈ λ(p_{ν(i)});
                   let r_a be the relation associated to σ_b(a) in DB;
                   if |π_{var_o(a)}(s[i]) ⋈ r_a|/|r_a| > k_sup then return true;
               end;
               return false;
           end; { ENOUGHSUPPORT }
       begin
           if ENOUGHSUPPORT then begin
               compute from s the relation b = J(σ_b(body(MQ)));
               for each type T instantiation σ_h for head(MQ) that agree with σ_b do begin
                   Set h to be the relation associated to head(MQ) by σ_h;
                   h' := h ⋈ b;
                   if |h'|/|h| > k_cvr and |b ⋈ h'|/|b| > k_cnf then Σ := Σ ∪ {σ_h ∘ σ_b};
               end
           end
       end; { FINDHEADS }
       procedure FINDBODIES(i : integer; σ_b : an instantiation);
       begin
           if i ≤ n then (* first half *)
               for each type T instantiation σ_i for λ(p_{ν(i)}) that agree with σ_b do begin
                   r[i] := π_{χ(p_{ν(i)})}(J(σ_i(λ(p_{ν(i)}))));
                   for each child p_j of p_{ν(i)} do r[i] := r[i] ⋈ r[ν^{-1}(j)];
                   if r[i] ≠ ∅ then FINDBODIES(i + 1, σ_b ∘ σ_i);
               end
           else begin (* second half *)
               s[n] := r[n];
               for j := n − 1 downto 1 do s[j] := r[j] ⋈ s[ν(father(ν^{-1}(j)))];
               FINDHEADS(σ_b);
           end;
       end; { FINDBODIES }
begin
   compute c, the hypertree-width of body(MQ);
   compute ⟨T, χ, λ⟩, a complete hypertree decomposition for body(MQ) of width c;
   compute a bottom-up visit of T = ({p_1, ..., p_n}, E) and encode it as a permutation ν of {1, ..., n};
   Σ := ∅;
   FINDBODIES(1, ∅);
   return Σ;
end { FINDRULES };
```

Figure 4: Algorithm FindRules

**Proposition 4.9** Let **MQ** be a metaquery and **DB** a database. Let $\langle \mathcal{T}, \chi, \lambda \rangle$ be a hypertree decomposition of width $c$ for **MQ**, and let $\sigma_{\mathbf{DB}}^{\mathbf{MQ}}$ be an instantiation. Define and $\lambda'(p) = \sigma(\lambda(p))$ for each $p \in vertices(\mathcal{T})$. Then $\langle \mathcal{T}, \chi, \lambda' \rangle$ is an hypertree decomposition of width $c$.

**Example 4.10** The width of $\langle \mathcal{T}^{ex}, \chi^{ex}, \lambda^{ex} \rangle$ of Example 4.8 is 2. As $Q^{ex}$ is not semi-acyclic, it follows that 2 is also the hypertree-width of $Q^{ex}$.

Let $Q$ be a set of atoms on a database **DB**, and let $\langle \mathcal{T}, \chi, \lambda \rangle$ be a hypertree decomposition of $Q$. It is known that we can build from $Q$, **DB** and $\langle \mathcal{T}, \chi, \lambda \rangle$, a set of atoms $Q'$, a database $\mathbf{DB}'$, and a join tree $\mathcal{T}'$ for $Q'$, such that $Q'$ is semi-acyclic and $\mathbf{J}(Q)$ on **DB** is equal to $\mathbf{J}(Q')$ on $\mathbf{DB}'$ [17]. Denote the above triplet $\langle Q', \mathbf{DB}', \mathcal{T}' \rangle$ as $acy(Q, \mathbf{DB}, \langle \mathcal{T}, \chi, \lambda \rangle)$. $\mathcal{T}'$ has the same tree shape as $\mathcal{T}$. For each vertex $p$ of $\mathcal{T}$, there is precisely one vertex $p'$ in $\mathcal{T}'$, and one relation $r'$ in $\mathbf{DB}'$. $p'$ is an atom having $\chi(p)$ as arguments and $r'$ is set to $\pi_{\chi(p)}(\mathbf{J}(\lambda(p)))$. $Q'$ contains all the atoms corresponding to vertices of $\mathcal{T}'$.

**Example 4.11** Consider the set of atoms $Q^{ex}$ of Example 4.8, the associated hypertree decomposition $\langle \mathcal{T}^{ex}, \chi^{ex}, \lambda^{ex} \rangle$, and the database $\mathbf{DB}^{ex} = \{p, q, r, s\}$. Let $p'_1 = p(A, B)$, $p'_2 = q(B, C)$, and $p'_3 = t(B, C, D) = r(C, D) \bowtie s(B, D)$, let $\mathcal{T}'^{ex}$ be the tree $\langle \{p'_1, p'_2, p'_3\}, \{\langle p'_1, p'_2 \rangle, \langle p'_2, p'_3 \rangle\} \rangle$, let $Q'^{ex} = \{p(A,B), q(B,C), t(B,C,D)\}$, and let $\mathbf{DB}'^{ex} = \{p, q, t\}$. Then $acy(Q^{ex}, \mathbf{DB}^{ex}, \langle \mathcal{T}^{ex}, \chi^{ex}, \lambda^{ex} \rangle) = \langle Q'^{ex}, \mathbf{DB}'^{ex}, \mathcal{T}'^{ex} \rangle$.

**Theorem 4.12** Let **r** be a Horn rule and let **DB** be a database, and let $d$ be the size of the largest relation within **DB**. Then $sup(\mathbf{r})$ can be computed in time $d^c \log d$, where $c$ is the hypertree-width of $body(\mathbf{r})$.

**Proof.** We can compute $sup(\mathbf{r})$ as follows:

1. set $Q = body(\mathbf{r}) = r_1(\mathbf{X}_1), \ldots, r_m(\mathbf{X}_m)$;

2. compute the hypertree width $c$ of $Q$ and a complete hypertree decomposition $\langle \mathcal{T}, \chi, \lambda \rangle$ of $Q$ having width $c$;

3. compute $acy(Q, \mathbf{DB}, \langle \mathcal{T}, \chi, \lambda \rangle) = \langle Q', \mathbf{DB}', \mathcal{T}' \rangle$;

4. compute the database $\overline{\mathbf{DB}'}$, formed by the reduced set of relations $Q'$ of $\mathbf{DB}'$, by executing a full reducer for $Q'$ (note that $Q'$ is semi-acyclic);

5. compute from $\overline{\mathbf{DB}'}$ the database $\overline{\mathbf{DB}}$ containing the reduced set $Q'' = \{r''_1(\mathbf{X}_1), \ldots, r''_m(\mathbf{X}_m)\}$ of relations in $Q$. (see [17] for details);

6. compute the sizes $d_1, \ldots, d_m$ of the relations in $Q$ and the sizes $d'_1, \ldots, d'_m$ of the relations in $Q''$;

7. compute the value of $sup(\mathbf{r})$ as $\max_i(d'_i/d_i)$.

Computing the hypertree-width $c$ of $Q$ and a $c$-width hypertree decomposition of $Q$ requires a constant amount of time in the data complexity measure. It also follows that the size of $\langle \mathcal{T}, \chi, \lambda \rangle$ is a constant. Therefore in the third step the only operation depending on the input size is the computation of $\pi_{\chi(p)}(\mathbf{J}(\lambda(p)))$, for each vertex $p$ of $\mathcal{T}$. Being $c$ the hypertree-width of $Q$ this can be done in time $d^c$. An upper bound for the time complexity of steps 4, 5, and 6 is $d^c \log d^c$. Finally, also step 7 requires a constant amount of time. □

**Definition 4.13** Let $S$ be a set of literal schemes. We denote by $pv(S)$ the set of predicate variables occurring in $S$. Let $\sigma_{\mathbf{DB}}^{S_1}$ and $\mu_{\mathbf{DB}}^{S_2}$ be two instantiations defined on two set of literal schemes $S_1$ and $S_2$ over a database $\mathbf{DB}$, respectively. We say that $\sigma$ and $\mu$ *agree* if

1. $\sigma(S) = \mu(S)$, where $S = S_1 \cap S_2$, and

2. $\sigma'(V) = \mu'(V)$, where,

   - $\sigma'$ and $\mu'$ are the restrictions of $\sigma$ and $\mu$ on the sets $pv(S_1)$ and $pv(S_2)$ of predicate variables of $S_1$ and $S_2$, respectively, and
   - $V = pv(S_1) \cap pv(S_2)$.

Clearly, if $\sigma$ and $\mu$ agree, then $\sigma \circ \mu$ is an instantiation on the set $S_1 \cup S_2$.

Figure 4 shows Algorithm FINDRULES that, given in input a database $\mathbf{DB}$, a metaquery $\mathbf{MQ}$, three rational numbers $0 \leq k_{sup}, k_{cvr}, k_{cnf} < 1$, and $T \in \{0, 1, 2\}$, computes all the type-$T$ instantiations $\sigma_{\mathbf{DB}}^{\mathbf{MQ}}$ such that $sup(\sigma(\mathbf{MQ})) > k_{sup}$, $cvr(\sigma(\mathbf{MQ})) > k_{cvr}$, and $cnf(\sigma(\mathbf{MQ})) > k_{cnf}$.

Within the main procedure, the algorithm computes the hypertree decomposition $\langle \mathcal{T}, \chi, \lambda \rangle$ of $body(\mathbf{MQ})$. By Proposition 4.9, this decomposition can be employed for each instantiation generated. Then, the procedure FINDBODIES performs a bottom-up visit of $\mathcal{T}$ that generates the instantiations $\sigma_b$ of the body of the metaquery by composing the instantiations $\sigma_i$ of the visited literal schemes. Also it computes the databases $\mathbf{DB}'$ and $\overline{\mathbf{DB}'}$ (see the proof of Theorem 4.12, steps (3) and (4)), and mantains in a data structure (an array $r$) the portions of the database $\overline{\mathbf{DB}'}$ shared by the partial instantiations generated during the visit, reusing them. When the root of the tree $\mathcal{T}$ is attained, the first half of a full reducer for $\sigma_b(body(\mathbf{MQ}))$ has been calculated. Hence the procedure FINDBODIES computes the second half of this full reducer (storing the reduced relations into an array $s$) and calls the procedure FINDHEADS. A backtracking on previously generated local substitutions is finally performed.

The procedure FINDHEADS then verifies if $\sigma_b(body(\mathbf{MQ}))$ has large enough support (this is done by calling the function ENOUGHSUPPORT) and, in the positive case, it computes the associated relation, and searches for heads such that the instantiated metaquery has large enough cover and confidence.

As for the time-complexity analysis of algorithm FINDRULES, in the data complexity setting, we note that during the bottom-up visit of $\mathcal{T}$ it performs a semijoin step for each edge of the visited tree, and that the number of edges visited by the algorithm is upper bounded by the number of instantiations of the body of the metaquery.

Then we can compute all the instantiations $\sigma_{\mathbf{DB}}^{\mathbf{MQ}}$ such that $sup(\sigma(\mathbf{MQ})) > k_{sup}$ in $n^{m-1}d^c \log d$ steps for $T \in \{0, 1\}$, and in $(nb^a)^{m-1}d^c \log d$ steps for $T = 2$, where $n$ is the number of relations within $\mathbf{DB}$, $m$ the number of relation patterns of the metaquery $\mathbf{MQ}$, $a$ the maximum arity of any relation pattern in $\mathbf{MQ}$, $b$ the maximum arity of any relation in $\mathbf{DB}$, and $c$ is the hypertree-width of $body(\mathbf{MQ})$. Note that $a$, $m$ and $1 \leq c \leq m - 1$ are constants. Moreover $ma$ is an upper bound for the metaquery size, $nbd$ is an upper bound for the number of attribute values in the database $\mathbf{DB}$, and usually holds that $d \gg nb$.

The additional steps needed to search for instantiations with high cover and confidence requires a total amount of $(nd)^m$ steps for $T \in \{0, 1\}$, and $(nb^a d)^m$ steps for $T = 2$.

## 5 Conclusions

In this paper we have formally defined the semantics of metaqueries and analyzed their computational complexity. In general, as far as the combined complexity is concerned, metaquerying is intractable (unless $P=NP$). Therefore, we have defined the class of acyclic metaqueries and shown that a subset of metaquerying problems defined by acyclic metaqueries is in $LOGCFL$ and, as such, is highly parallelizable. Moreover, we have studied the data complexity of metaquerying problems and showed that, in general, it lies within $TC^0$. Even in this case, we were able to single out a subset of metaquerying problems, for which the data complexity is in $AC^0$. Finally, we have discussed metaquery implementation. The complexity analysis of metaquerying problems presented here, and which is summarized in Figure 5, is not complete, though. We are working towards establishing other results. In particular, it seems to be interesting to extend our formal framework to more general metaquery forms, such as allowing negation and/or disjunction to occur in metapatterns.

| Complexity measure | Problem type | Instantiation type | Indices | Threshold | Complexity |
|---|---|---|---|---|---|
| Combined Complexity | General | 0, 1, 2 | **I** | $k = 0$ | NP-Complete (Th. 3.21) |
| Combined Complexity | General | 0, 1, 2 | $cvr$ and $sup$ | $0 \leq k < 1$ | NP-Complete (Th. 3.24) |
| Combined Complexity | General | 0, 1, 2 | $cnf$ | $0 \leq k < 1$ | $NP^{PP}$-Complete (Th. 3.28, 3.29) |
| Combined Complexity | Acyclic | 0 | **I** | $k = 0$ | LOGCFL-Complete (Th. 3.32) |
| Combined Complexity | Acyclic | 1, 2 | **I** | $k = 0$ | NP-Complete (Th. 3.33) |
| Combined Complexity | Acyclic | 0 | $cvr$ and $sup$ | $0 \leq k < 1$ | Open |
| Combined Complexity | Acyclic | 1, 2 | $cvr$ and $sup$ | $0 \leq k < 1$ | NP-complete (Th. 3.34) |
| Combined Complexity | Acyclic | 0, 1, 2 | $cnf$ | $0 \leq k < 1$ | Open |
| Combined Complexity | Semi-acyclic | 0, 1, 2 | **I** | $k = 0$ | NP-Complete (Th. 3.35,Th. 3.33) |
| Data Complexity | General | 0, 1, 2 | **I** | $k = 0$ | $AC_0$ (Th. 3.37) |
| Data Complexity | General | 0, 1, 2 | **I** | $0 \leq k < 1$ | $TC_0$ (Th. 3.38) |

Figure 5: Summary of complexity results.

# 6  Acknowledgements

The authors thank Francesco Scarcello for several interesting discussions. Special thanks to Arnaud Durand, who provided illuminating insights about some of the computational complexity analysis tools we have used in this paper.